\documentclass[a4paper,12pt]{article}

\pdfoutput=1

\makeatletter

\@addtoreset{equation}{section}
\makeatother
\usepackage{amsfonts}
\usepackage{amsmath}

\usepackage{graphicx}

\def\be{\begin{equation}}
\def\ee{\end{equation}}
\def\bea{\begin{eqnarray}}
\def\eea{\end{eqnarray}}
\def\({\left(}
\def\){\right)}
\def\<{\left<}
\def\>{\right>}

\def\[{\left[}
\def\]{\right]}

\def\be{\begin{equation}}
\def\ee{\end{equation}}
\def\bea{\begin{eqnarray}}
\def\eea{\end{eqnarray}}

\def\({\left(}
\def\){\right)}
\def\<{\left<}
\def\>{\right>}

\def\tr{{\mbox{tr}}}
\def\be{\begin{equation}}
\def\ee{\end{equation}}
\def\bea{\begin{eqnarray*}}
\def\eea{\end{eqnarray*}}
\def\ben{\begin{eqnarray}}
\def\een{\end{eqnarray}}
\def\({\left(}
\def\){\right)}
\def\<{\left<}
\def\>{\right>}
\def\!{\right|}
\def\|{\left|}

\def\[{\left[}
\def\]{\right]}

\def\+{\bar}
\def\mb{\mathbb}
\def\tr{{\mbox{tr}}}

\def\Vol{{\mbox{Vol}}}

\def\L{{\cal{L}}}
\def\t{\widetilde}
\def\A{{\cal{A}}}

\def\M{{\cal{M}}}

\def\O{{\cal{O}}}

\def\L{{\cal{L}}}

\def\L{{\cal{L}}}

\def\eps{{\cal{\varepsilon}}}

\def\l{{{\ell}}}

\begin{document}
\pagestyle{empty}
\vskip-10pt
\vskip-10pt
\begin{center}
\vskip 3truecm
{\Large\bf
Renormalization of the Einstein-Hilbert action}
\vskip 2truecm
{\large \bf
Andreas Gustavsson}
\vspace{1cm} 
\begin{center} 
Physics Department, University of Seoul, 13 Siripdae, Seoul 130-743 Korea
\end{center}
\vskip 0.7truecm
\begin{center}
(\tt agbrev@gmail.com)
\end{center}
\end{center}
\vskip 2truecm
{\abstract We examine how the Einstein-Hilbert action is renormalized by adding the usual counterterms and additional corner counterterms when the boundary surface has corners. A bulk geometry asymptotic to $H^{d+1}$ can have boundaries $S^k \times H^{d-k}$ and corners for $0\leq k<d$. We show that the conformal anomaly when $d$ is even is independent of $k$. When $d$ is odd the renormalized action is a finite term that we show is independent of $k$ when $k$ is also odd. When $k$ is even we were unable to extract the finite term using the counterterm method and we address this problem using instead the Kounterterm method. We also compute the mass of a two-charged black hole in AdS$_7$ and show that background subtraction agrees with counterterm renormalization only if we use the infinite series expansion for the counterterm.}

\vfill
\vskip4pt
\eject
\pagestyle{plain}

\section{Introduction}
The AdS-CFT correspondence relates Einstein gravity in the bulk with a conformal field theory on the boundary. A deformation of the boundary gives rise to a conformal transformation of the induced metric on the boundary. The shape of the boundary submanifold that is placed near infinity is not supposed to affect the gravity action very much. However, if we impose a Dirichlet boundary condition at the boundary, then the on-shell value of the gravity action becomes a function of the boundary metric. At this stage the on-shell value depends heavily on the choice of boundary through the boundary metric, which is a crucial observation for the Hamilton-Jacobi theory and the Brown-York quasilocal stress tensor \cite{Brown:1992br}. However, if the boundary is viewed as a regulator surface near infinity, then we may subtract the divergent terms in the on-shell action by adding a counterterm \cite{Emparan:1999pm}, \cite{Balasubramanian:1999re}, \cite{Kraus:1999di}. This is an action on the boundary that makes use only of the intrinsic geometry of the boundary. Adding the counterterm does not affect the bulk gravity equations of motion that are derived by keeping the boundary metric fixed. By adding the counterterm action we get a renormalized gravity action that has no divergences as the boundary is taken towards infinity. We expect that this renormalized action is insensitive to the precise location and shape of the boundary manifold, up to correction terms that go to zero as the boundary moves to infinity. 

Classical gravity breaks down at a singularity in the bulk. On the other hand we expect that a singularity in the boundary geometry, which is caused by the embedding in a smooth bulk geometry, should be completely harmless with no physically observable consequences. The boundaries that we will study in this paper will always be peacewise smooth and joined at corners and we will for the most part restrict ourselves to Euclidean spacetimes. There can be other types of singularities in the boundary such as conical singularities \cite{Fursaev:1995ef}. If one assumes the spacetime is Minkowskian, one needs to distuish between timelike, spacelike and lightlike boundaries \cite{Jubb:2016qzt}, \cite{Parattu:2016trq}, \cite{Lehner:2016vdi}, \cite{Jafari:2019bpw}. The null boundary can be the horizon of a blackhole \cite{Chakraborty:2019doh}. However, in this paper we will keep it simple and consider Euclidean spacetime where we treat the boundary as a regulator surface near infinity of AdS. Viewed as a regulator it is the usual story that the corresponding renormalized quantity should not depend on the choice of regulator. In this paper we will show by some examples, at least partially but still rather convincingly, that the renormalized value of the gravity action is independent of the shape of the boundary, and that this remains true also in the presence of corners.

There are two important classes of boundaries that one needs to distinguish, namely when the dimension $d$ of the boundary is even and odd respectively. If $d$ is even, then there is a conformal anomaly $\A$ \cite{Henningson:1998gx} that should be invariant under a deformation of the boundary. If $d$ is odd, the renormalized gravity action is a finite constant $F$ \cite{Anastasiou:2019ldc}, \cite{Pufu:2016zxm}, \cite{Taylor:2016kic} that should be invariant under a deformation of the boundary. 

The gravity action in Euclidean signature with a negative cosmological constant $\Lambda$ reads
\bea
I_{bulk} &=& - \frac{1}{16 \pi G} \int d^{d+1} x \sqrt{g} \(R - 2 \Lambda\)
\eea
To this action one adds a surface terms \cite{Gibbons:1976ue}
\bea
I_{surf} &=& - \frac{1}{8\pi G} \int d^d x \sqrt{h} K 
\eea
where $K$ is the trace of the extrinsic curvature tensors computed with an outward pointing unit normal vector and $h_{\mu\nu}$ is the induced boundary metric. If we vary the bulk metric we find that the variation of $I = I_{bulk} + I_{surf}$ is proportional to the variations of the boundary metrics. By keeping the boundary metric fixed we derive Einstein's equations of motion in the bulk. The on-shell value is a function of the boundary metric, $I = I(h_{\mu\nu})$. To this action, one may add a boundary term that only depends on the boundary metric and its tangential derivatives without affecting the bulk equations of motion. This can be used to construct a counterterm action \cite{Emparan:1999pm}, \cite{Balasubramanian:1999re}, \cite{Kraus:1999di}\footnote{In the last term we show in this expansion there are additional derivative terms that we do not display as they will not contribute to the particularly symmetric boundaries that we consider in this paper.}
\ben
I_{ct} &=& \frac{1}{8\pi G} \int d^d x \sqrt{h} \L_{ct}\cr
\L_{ct} &=& \frac{d-1}{\l} + \frac{\l}{2(d-2)} R + \frac{\l^3}{2(d-2)^2(d-4)} \(R_{ij}^2 - \frac{d}{4(d-1)} R^2\)\cr
&& - \frac{\l^5}{(d-2)^3(d-4)(d-6)} \(\frac{3d+2}{4(d-1)} R R_{\mu\nu}^2 - \frac{d(d+2)}{16(d-1)^2}R^3 - 2 R_{\mu\nu\kappa\tau} R^{\mu\kappa} R^{\nu\tau}\) \cr
&& + ...\label{ct}
\een
that cancels the powerlaw divergent terms as the boundary is taken to infinity and we define the renormalized action as $I_{ren} = I_{bulk} + I_{surf} + I_{ct}$. 

We notice that there are poles such as $1/(d-2)$ and $1/(d-4)$ and so on in the counterterm at every even dimension. So when $d$ is even, one may have to truncate the counterterm series expansion at the term before one hits such a pole singularity \cite{Emparan:1999pm}. But as we will see, there are exceptions when the boundary has particular high degree of symmetry and these pole singularities are canceled by the curvature invariants. In such cases we shall not truncate the counterterm.

Also in odd dimensions, there is no reason why we shall truncate the counterterm. But often we may do it. But this gives the right answer only for cases when there are no divergences in the terms one truncates. Such divergences can arise when the boundary is noncompact. We will see examples when we can not truncate the counterterm for odd $d$ due to such  divergences. 

The organization of this paper is as follows. In section \ref{boundary} we focus on trying to understand what happens when the boundary is noncompact and needs to be regularized, thus  introducing a boundary on the boundary\footnote{There is no boundary of a boundary. But it may serve as an intuitive phrase. What we really get are boundary segments joined at corners.}. In our case the boundary is $S^k \times H^{d-k}$ where $H^{d-k}$ is noncompact and needs to be regularized. This is a difficult problem. We are able to solve it completely only when the boundary is one-dimensional. In higher dimensions we get many interesting partial results and find evidence that the finite term in the renormalized action is universal when $d$ is odd. In section \ref{conformal} we turn to the problem of extracting the logarithmic divergence and the conformal anomaly on $S^k \times H^{d-k}$ when $d$ is even. Perhaps not too surprisingly, we find that the conformal anomaly does not depend on $k$. But the cutoff dependence for the logarithm is completely different depending on whether $k$ is even or odd. When $k$ is even, the log dependence is just the usual one of the cutoff boundary surface. But when $k$ is odd, the log dependence is of the cutoff of the boundary of the boundary. We also obtain the exact form of the counterterm on the boundary and we find pole cancelations that for even dimensional boundaries mean that the counterterm shall not be truncated, as one usually does. We show by an explicit example that truncation gives the wrong conformal anomaly. In section \ref{Kounterterm} we compare the counterterm renormalization with the Kounterterm renormalization and find that the Kounterterm works nicely for all the cases we checked. In section \ref{mass} we apply the counterterm renormalization to compute the mass of AdS and the mass of a black hole. We find agreement with the background subtraction method, and this provides further evidence that the counterterm shall not be truncated. In section \ref{formula} we propose another formula for the counterterm that is inspired by the Mann-Marolf counterterm in flat space. We also obtain a $1/d$ expansion of the counterterm in the flat space limit. There are two appendices. In appendix B we compute an integral that enter in all computations in this paper. There are two figures that we placed at the end of the paper.

\section{Noncompact boundaries}\label{boundary}
We would like to understand when the conformal anomaly can be read off from the coefficient of the logarithmic divergence of a cutoff that goes to zero as we take the boundary to infinity. On the hyperbolic space $H^{d+1}$ we can put the metric 
\ben
ds^2 &=& d\rho^2 + \sinh^2\rho d\Omega_d^2\label{spherefoliation}
\een
where $d\Omega_d^2$ is the metric on unit $S^d$. This metric foliates $H^{d+1}$ by spheres $S^d$, which are level surfaces at constant $\rho\in [0,\infty]$. A Fefferman-Graham coordinate system is $(u,\varphi)$ where 
\ben
u &=& e^{-\rho}\label{u}
\een
The metric becomes
\bea
ds^2 &=& \frac{1}{u^2} \(du^2 + \frac{1}{4}\(1-u^2\)^2 d\Omega_d^2\)
\eea
and this metric is on the FG form. Once the metric is on this FG form, the metric in the conformal field theory that lives on the boundary of AdS may be defined as \cite{Witten:1998qj}
\bea
dS^2 = \lim_{u\rightarrow 0} \(f(u)^2 ds^2\)
\eea
where $f(u)$ is a positive function in the bulk and has a first order zero on the boundary. This insures that the CFT metric $dS^2$ on the boundary is well-defined and finite. Given one such a defining function $f$, any other such function is related to it by $f \rightarrow e^{\Omega} f$ where $\Omega$ is any function on the entire space, the bulk plus boundary. This transformation induces a conformal transformation on the CFT metric. This also shows that the CFT metric of two different FG coordinates are related by a conformal transformation in the following sense. If we use $u$ as defined by (\ref{u}) as our FG coordinate in $H^{d+1}$ and place the boundary at $u = \eps$, then we get the CFT metric
\bea
dS^2_{u=\eps} &=& \frac{1}{4} d\Omega_d^2 + \O(\eps)
\eea
If we use $v = e^{-\Omega} u$ as our FG coordinate and place our boundary at $v=\eps$, then we get the CFT metric
\bea
dS^2_{v=\eps} &=& e^{-2\Omega} \frac{1}{4} d\Omega_d^2 + \O(\eps)
\eea
These two metrics are not describing the same boundary manifold since the cutoff surfaces that define them are embedded in $H^{d+1}$ in different ways. As we take $\eps\rightarrow 0$, the two CFT metrics are related to one another by a conformal transformation.

If we use an FG coordinate $u$ near the boundary such that the boundary is located at $u=\eps$, then the gravity action $I = I_{bulk} + I_{surf}$ takes the form \cite{Henningson:1998gx}
\bea
I &=& \frac{a_0}{\eps^d} + \frac{a_{1}}{\eps^{d-2}} + \cdots + \frac{a_{(d-1)/2}}{\eps} + F + \O(\eps)\cr
I &=& \frac{a_0}{\eps^d} + \frac{a_{1}}{\eps^{d-2}} + \cdots + \frac{a_{d/2-1}}{\eps^2} + \A \ln\eps + a_{d/2+1} + \O(\eps)
\eea
for odd and even $d$ respectively. Here $\A$, which is the coefficient of the log divergent term, can be identified as the conformal anomaly precisely because $\eps$ is a cutoff of an FG coordinate. The conformal anomaly is known to be conformally invariant. On the other, $F$ is just a finite term. Is this also conformally invariant? To remove the powerlaw divergences we need to add a counterterm action $I_{ct}$ that is constructed out of the intrinsic geometry of the boundary. It could happen that the counterterm could have an ambiguous finite term $F_{ct}$. If so, then $F_{ren} = F+F_{ct}$ could become ambiguous. Yet we believe that $F_{ren}$ is not ambigious. We will present examples where the combination $F_{ren}$ remains the same for several different choices of the conformal boundary.

It remains to construct the renormalized action. When the boundary is smooth, it takes the form
\bea
I_{ren} &=& I_{bulk} + I_{surf} + I_{ct}
\eea
When there are corners on the boundary, the form of the renormalized action is unknown. What is known is that in order for the unrenormalized gravity action to have a well-defined variational principle, we need to add a certain corner term $I_{corn}$ that has been studied by many authors in various context. For a sample of literature, see for instance \cite{Jubb:2016qzt}, \cite{Parattu:2016trq}, \cite{Lehner:2016vdi}, \cite{Jafari:2019bpw}. But to renormalize the action we need to add many more terms.

We will now examine possible FG coordinates in $H^2$ that we embed into $\mb{R}^{1,2}$ with the metric 
\bea
ds^2 = - (dX^0)^2 + (dX^1)^2 + (dX^2)^2
\eea
as the hypersurface 
\bea
-(X^0)^2 + (X^1)^2 + (X^2)^2 &=& -1
\eea
There are two branches with $X^0\geq 1$ and $X^0\leq -1$. In this paper we will focus on the branch $X^0\geq 1$. We can parametrize $H^2$ in two different ways as
\begin{alignat}{3}
X^0 &= \cosh\rho &&= \cosh\t\rho\cosh\eta\label{ett}\\
X^1 &= \sinh\rho \cos\varphi &&= \cosh\t\rho\sinh\eta\label{tva}\\
X^2 &= \sinh\rho \sin\varphi &&= \sinh\t\rho\label{tre}
\end{alignat}
that gives the two different metrics
\ben
ds^2 &=& d\rho^2 + \sinh^2\rho d\varphi^2\label{fol1}\\
ds^2 &=& d\t\rho^2 + \cosh^2\t\rho d\eta^2\label{fol2}
\een
The former foliation has level surfaces the are circles $(X^1,X^2) = r (\cos\varphi,\sin\varphi)$ with radius $r=\sinh\rho\geq 0$ at constant $X^0 = \sqrt{r^2 + 1}$. The latter foliation has level surfaces that are hyperbolas $(X^0,X^1) = r \(\cosh\eta,\sinh\eta\)$ for $-\infty < \eta < \infty$ at constant $X^2 = \pm \sqrt{r^2-1}$ for $r = \cosh\t\rho \geq 1$ and therefore the boundary at some constant large $r$ is noncompact and needs to be regularized.

We illustrate these two foliations of $H^2$ as a surface in $\mb{R}^{1,2}$ in Figures 1 and 2 on the last page of this paper. For the hyperbolic foliation (\ref{fol2}), the regularized boundary consists of four boundary segments that are joined at corners with a 90 degrees deficit angle. These angles appear to be much sharper than $90$ degrees in the figure. This is  because as the corners approach the lightcone in $\mb{R}^{1,2}$ they are stretched out. 

For the circle foliation (\ref{fol1}) we have $\varphi \in [0,2\pi]$ and $\rho \in [0,\rho_0]$ where $\rho_0 = \ln(1/\eps)$ is a large cutoff. The boundary at $\rho = \rho_0$ has the boundary metric\footnote{We will distinguish between boundary metric and CFT metric. The boundary metric is the metric that is induced from the bulk the usual way, and this is divergent in the limit $\eps\rightarrow 0$.}
\bea
ds^2 &=& \frac{1}{4\eps^2} d\varphi^2  
\eea
This behavior of the boundary metric makes it clear that a constant rescaling of $\eps$ induces a corresponding constant rescaling of the metric, which suggests that we can read off the conformal anomaly from the coefficient of $\log (1/\eps)$. 
 
Let us now consider the second metric (\ref{fol2}). For (\ref{tre}) to generalize to higher dimensions we should  write $X^2 = \pm \sinh\t\rho$ and let $\t\rho \in [0,\t\rho_0]$ reflecting the fact that $S^0 = \{\pm 1\}$. Here we do not write the $\pm$, so instead we must take $\t\rho\in [-\t\rho_0,\t\rho_0]$. The most natural choice of boundary for this second representation of the metric has four boundary segments 
\bea
B_{I\pm} &=& \{\t\rho = \pm\t\rho_0, \eta\in [-\eta_0,\eta_0]\}\cr
B_{II\pm} &=& \{\t\rho \in [0,\t\rho_0], \eta = \pm\eta_0\}
\eea
joined at four corners $(\rho,\eta) = \{(\rho_0,\eta_0), (\rho_0,-\eta_0),(-\rho_0,\eta_0),(-\rho_0,-\eta_0)\}$. The boundary metric on each boundary segment is
\ben
ds^2_{I\pm} &=& \cosh^2 \t\rho_0 d\eta^2\label{toxe}\\
ds^2_{II\pm} &=& d\t\rho^2\label{exot}
\een
On the boundary segments $B_{I\pm}$, things are still quite familiar. We may define a coordinate $v = e^{-\t\rho}$ and a cutoff $\t\eps = e^{-\t\rho_0}$ and the boundary metric becomes
\bea
ds^2_{I\pm} &=& \frac{1}{4 \t\eps^2} d\eta^2
\eea
This shows that we can read off the conformal anomaly as the coefficient of $\log\(1/\t\eps\)$. However, for the boundary segments $B_{II\pm}$ it is less obvious that we can read off the conformal anomaly as the coefficient of $\log\(1/\delta\)$ where $\delta = e^{-\eta_0}$. The origin to this problem should lie in the bad choice of coordinates that are not of FG form. However, $(u,\varphi)$ are FG coordinates. This motivates us to re-express the boundary metric in terms of the coordinate $\varphi$ using the relation
\bea
\tanh\t\rho &=& \tan\varphi \sinh\eta_0
\eea
Here $\varphi$ ranges as follows for the various boundary segments,
\bea
B_{I+} &=& \{\varphi \in [\varphi_0,\pi/2]\}\cr
B_{I-} &=& \{\varphi \in [-\pi/2,-\varphi_0]\}\cr
B_{II+} &=& \{\varphi \in [0,\varphi_0]\}\cr
B_{II-} &=& \{\varphi \in [-\varphi_0,0]\}
\eea
For the boundary segments $B_{II\pm}$, we get
\bea
ds = d\t\rho = \frac{\sinh\eta_0}{\cos^2\varphi-\sinh^2\eta_0\sin^2\varphi} d\varphi
\eea
This is essentially on the FG form
\bea
ds &=& \frac{1}{\delta} f(\varphi,\delta) d\varphi
\eea
with
\bea
f(\varphi,\delta) &=& \frac{1}{\cos^2\varphi - \sinh^2\eta_0 \sin^2\varphi} 
\eea
But we need to restrict ourselves to a very short interval, say $0\leq \varphi \leq \delta^{3/2}$, where we stay far away from the corner singularity. On this very short interval, we have the desired FG behavior of the metric with $f(\varphi,\delta) = f(\varphi) + \O(\delta)$. Since conformal transformations can act locally, this should be sufficient for us to read off the conformal anomaly from the coefficient of $\ln(1/\delta)$ in the renormalized action.

This has a generalization to higher dimensional hyperbolic spaces. On $H^{d+1}$ we have a family of metrics 
\ben
ds^2 &=& d\rho^2 + \sinh^2 \rho d\Omega_k^2 + \cosh^2\rho \(d\eta^2 + \sinh^2\eta d\Omega_{d-k-1}^2\) \label{folk}
\een
for $k=0,1,...,{d-1}$ where $d\Omega_k^2$ is the metric on unit $S^k$. When $k=d$ we have the metric 
\bea
ds^2 &=& d\rho^2 + \sinh^2\rho d\Omega_d^2
\eea
The conformal anomaly can be extracted from the coefficient of either $\eta_0 = \ln(1/\delta)$ or $\rho_0 = \ln(1/\eps)$ when we expand the on-shell gravity action.

We will now proceed to compute the renormalized action for $H^{d+1}$ where we use the foliation (\ref{folk}). The bulk action and the surface terms can be expressed for general $d$ and $k$, while the counterterm and possible other terms associated with the corner depend on the dimension and are difficult to express for general $d$ and $k$. We have the following on-shell value of the bulk action for general $d$ and $k$,
\ben
I_{bulk,k} &=& \frac{d}{8\pi G} \Vol_k(H^{d+1})\cr
\Vol_k(H^{d+1}) &=& \Vol(S^k) \Vol(H^{d-k}) J_{d,k}\label{bulkaction}
\een
where we define the integral 
\bea
J_{d,k} &=& \int_0^{\rho_0} d\rho \sinh^k \rho \cosh^{d-k} \rho
\eea
that we compute in the appendix \ref{integral}. We denote by 
\bea
\Vol(S^d) &=& \frac{(d+1)\pi^{\frac{d+1}{2}}}{\Gamma\(\frac{d+3}{2}\)}
\eea
the volume of unit $S^d$ and by $\Vol(H^{d+1})$ the volume of $H^{d+1}$ when it is foliated as in (\ref{spherefoliation}). The surface term associated with the boundary $\rho = \rho_0$ is given by 
\bea
I^I_{surf,k} &=& - \frac{\Vol(S^k) \Vol(H^{d-k})}{8\pi G} \sinh^k \rho_0 \cosh^{d-k}\rho_0 \(k \frac{\cosh\rho_0}{\sinh\rho_0} + (d-k) \frac{\sinh\rho_0}{\cosh\rho_0}\)
\eea
The surface term associated with the boundary $\eta = \eta_0$ is given by 
\bea
I^{II}_{surf,k} &=& - \frac{\Vol(S^k) \Vol(S^{d-k-1})}{8\pi G} \(d-k-1\) \cosh\eta_0 \sinh^{d-k-2}\eta_0 J_{d-2,k} 
\eea
For both these surface terms we use the convention $\Vol(S^0) = 2$ and the coordinate ranges $\rho\in [0,\rho_0]$ and $\eta \in [0,\eta_0]$. 

We will also need the volumes of hyperbolic spaces when foliated as (\ref{spherefoliation}) and we regularize by taking $\eta\in [0,\eta_0]$. We then get the volume
\bea
\Vol(H^{d+1}) &=& \Vol(S^d) \int_0^{\eta_0} d\eta \sinh^d \eta
\eea
For the first few dimensions
\bea
\Vol(H^1) &=& 2\eta_0\cr
\Vol(H^2) &=& 2\pi \cosh\eta_0 - 2\pi\cr
\Vol(H^3) &=& \pi \sinh(2\eta_0) - 2\pi \eta_0
\eea
For odd $d$ we get the following finite term
\ben
\Vol_{ren}(H^{d+1}) &=& \pi^{\frac{d}{2}} \Gamma\(-\frac{d}{2}\)\label{renHodd}
\een
For even $d$ we get the following log divergence
\ben
\Vol_{ren}(H^{d+1}) &=& \frac{2\pi^{\frac{d}{2}} (-1)^{\frac{d}{2}}}{\Gamma\(\frac{d}{2}+1\)} \ln\(\frac{1}{\delta}\)\label{renHeven}
\een
where $\delta = e^{-\eta_0}$.

\subsection{One dimension}
For $d=1$ there is no conformal anomaly but there is a universal finite term instead. Let us show this explicitly for the two boundary surfaces that we constructed above. The on-shell bulk gravity action is 
\bea
I_{bulk} &=& \frac{1}{8\pi G} \Vol(H^2)
\eea
In the foliation (\ref{fol1}) we get
\bea
\Vol(H^2) &=& \Vol(S^1) \int_0^{\rho_0} d\rho \sinh \rho
\eea
and so 
\bea
I_{bulk} &=& \frac{1}{4G} \cosh\rho_0 - \frac{1}{4G}
\eea
The surface term is 
\bea
I_{surf} &=& - \frac{1}{4G} \cosh\rho_0
\eea
The counterterm is 
\bea
I_{ct} &=& 0
\eea
since $d-1 = 0$ and the intrinsic curvature in 1d is zero.  Summing all contributions, we get
\ben
I_{ren} = I_{bulk} + I_{surf} + I_{ct} = - \frac{1}{4G}\label{p}
\een
Let us now use the foliation (\ref{fol2}). Then we get
\bea
I_{bulk} &=& \frac{1}{2\pi G} \sinh \t\rho_0 \eta_0\cr
I_{surf} &=& - \frac{1}{2\pi G} \sinh \t\rho_0\eta_0\cr
I_{ct} &=& 0
\eea
A bit unexpectedly, we find log divergences $\sim \eta_0 = \ln(1/\delta)$ for the individual terms despite here $d=1$ is odd. However, when we add these contributions together we, these divergences cancel, which is very fortunate as we do not expect a conformal anomaly here, but we get $I_{bulk}+I_{surf}+I_{ct} = 0$ and this does not agree with the finite term that we got in (\ref{p}). Let us also notice that $I_{surf}^{II} = 0$ so the entire surface term contribution comes from the boundaries $B_{I\pm}$.

One could now argue that the finite term must depend on the renormalization scheme and is ambiguous. We could simply add a finite constant to the action as a counterterm. This would amount to shifting the potential energy. There are many reasons why we can not accept such a viewpoint. First, if we would write the counterterm as an integral over a Lagrangian density, we would need to multiply the shifted Lagrangian density by the length of the boundary, which diverges as we take the boundary to infinity. To make the counterterm action finite, the Lagrangian density would therefore have to be fine-tuned such that it is very close to zero as we take the boundary to infinity in such a way that the action stays finite. A dependence of the  Lagrangian density on the cutoff scale and the high degree of fine tuning is unnatural. The second reason is that we do not want to allow for a shift of the potential by an arbitrary constant. Although in many situations such a shift can not be detected experimentally, one exception occurs when one has supersymmetry. The third reason is that the counterterm can be obtained in a very general form that applies to any gravity theory in any dimension. For example, for pure gravity with a smooth boundary it is given by eq (\ref{ct}). We would like to propose that one shall always use such general form of the counterterm in every situation and never cook up something else that may seem to work particular cases. This last argument will be used again later and we will make it stronger as we compute the conformal anomaly later on in this paper.

There is an elegant way to get finite term right in this example. We have been ignorant about the fact that there are four corners on the boundary. These will contribute with finite corner terms. To see this, we may regularize a corner by replacing the sharp corner with an arc of a small circle with radius $r$. Since the corner is locally embedded in flat space (and this approximation becomes exact in the limit $r\rightarrow 0$), we may assume the ambient space is flat $\mb{R}^2$ instead of $H^2$ in the vicinity of a corner. The extrinsic curvature of a circle of radius $r$ embedded in flat space is given by  
\bea
K &=& \frac{1}{r}
\eea
The line element along the arc of the circle is $ds = r d\varphi$ where $\varphi$ parametrizes an angle ranging over say $\varphi \in [0,\alpha]$ where $\alpha$ is the angle that characterizes the corner (in our case $\alpha = \pi/2$). The Gibbons-Hawking surface term contribution that comes from this arc is now given by 
\bea
I_{arc} &=& - \frac{1}{8\pi G} r \int_0^{\alpha} d\varphi \frac{1}{r} = - \frac{\alpha}{8\pi G}
\eea
Since there are four corners along the entire boundary surface, we get in total 
\bea
I_{corn} &=& 4 \lim_{r\rightarrow 0} I_{arc} = - \frac{1}{4G}
\eea
where we put $\alpha = \pi/2$. Now we see that the correctly renormalized action becomes 
\begin{alignat*}{3}
I_{ren} &= I_{bulk} + I_{surf} + I_{ct} + I_{corn} &&= - \frac{1}{4G}
\end{alignat*}
in precise agreement with (\ref{p}). 

This is the only example where we have managed to get a precise cancelation of all divergences and also a perfect match of the finite term for two different boundaries. In higher dimensions, we believe that one can again cancel all divergences, but the problem becomes much more difficult and we have been able to only demonstrate a partial cancelation of divergences. However, again we have been able to match finite terms. 

Corner terms have been derived for boundaries of dimension $d\geq 2$ where they are given by 
\ben
I_{corn} &=& - \frac{\alpha}{8\pi G} \Vol(corn)\label{corner}
\een
We notice that $\alpha$ for a sharp corner has the interpretation as the deflection angle. We are not aware of a derivation of the corner term for a 1d boundary surface in the literature, but our 1d result follows from the general form of the corner term if we assign the volume of a point (which is the corner manifold when $d=1$) to be $\Vol(point) = 1$.

\subsection{Two dimensions}
Let us next consider $d=2$ and the bulk space $H^3$. Since $d=2$ is even, there will be a conformal anomaly. We know that the conformal anomaly is invariant under conformal transformations so it should not depend on $k$. We will here see that the same conformal anomaly appears for all values of $k$, which probably is an indication that we are on the right track. The simplest case is $k=2$ because then the cutoff boundary is $S^2$ and there are no corners on this boundary. If we define $e^{-\rho_0} = \eps$, then a standard computation gives the result
\bea
I_{ren} &=& - \frac{1}{2G} \ln \frac{1}{\eps}
\eea
We read off the conformal anomaly as the coefficient of the log-divergence. The precise way of defining the anomaly depends on a convention. We will use the simplest possible convention that the anomaly here is simply the coefficient of $\ln(1/\eps)$. Hence
\bea
\A &=& - \frac{1}{2G}
\eea
We would like to reproduce this conformal anomaly for the other cutoff boundaries corresponding to $k=0,1$. The bulk gravity action is 
\begin{align*}
I_{bulk,0} &= \frac{\Vol(S^0) \Vol(H^2)}{8\pi G} \(\rho_0 + \frac{1}{2} \sinh\(2\rho_0\)\)\cr
I_{bulk,1} &= \frac{\Vol(S^1) \Vol(H^1)}{8\pi G} \(\cosh^2\rho_0 - 1\)\cr
I_{bulk,2} &= \frac{\Vol(S^2) \Vol(H^0)}{8\pi G} \(-\rho_0 + \frac{1}{2}\sinh\(2\rho_0\)\)
\end{align*}
The surface term is 
\bea
I^I_{surf,0} &=& - \frac{\Vol(S^0) \Vol(H^2)}{8\pi G} \sinh\(2\rho_0\)\cr
I^I_{surf,1} &=& - \frac{\Vol(S^1) \Vol(H^1)}{8\pi G} \(\cosh^2\rho_0 + \sinh^2\rho_0\)\cr
I^I_{surf,2} &=& - \frac{\Vol(S^2) \Vol(H^0)}{8\pi G} \sinh\(2\rho_0\) 
\eea
The counterterm in $d=2$ is given by
\bea
I_{ct,k} &=& \frac{1}{8\pi G} \int d^2 x \sqrt{h}
\eea
so we get
\bea
I^I_{ct,0} &=& \frac{\Vol(S^0) \Vol(H^2)}{8\pi G} \cosh^2\rho_0\cr
I^I_{ct,1} &=& \frac{\Vol(S^1) \Vol(H^1)}{8\pi G} \frac{1}{2}\sinh\(2\rho_0\)\cr
I^I_{ct,2} &=& \frac{\Vol(S^2) \Vol(H^0)}{8\pi G} \sinh^2\rho_0
\eea
Adding these, and using the volumes 
\begin{align*}
\Vol(H^0) &= 1\cr
\Vol(H^1) &= 2 \ln\frac{1}{\delta}\cr
\Vol(H^2) &= \frac{\pi}{\delta} - 2\pi
\end{align*}
as derived in the appendix, we get
\bea
I^{naive}_{ren,0} &=& - \frac{1}{2G} \ln\frac{1}{\eps} + \frac{1}{4G} \frac{1}{\delta}\ln\frac{1}{\eps}\cr
I^{naive}_{ren,1} &=& - \frac{1}{4G} \ln\frac{1}{\delta}\cr
I^{naive}_{ren,2} &=& - \frac{1}{2G} \ln\frac{1}{\eps}
\eea
We see that the coefficient of $I_{ren,1}^{naive}$ is half of what we would expect. We will postpone the solution of that problem until section \ref{solved}.
 
The other surface term is
\bea
I^{II}_{surf,0} &=& - \frac{1}{2G} \rho_0 \cosh\eta_0\cr
I^{II}_{surf,1} &=& 0
\eea
and we see that the unwanted divergent term $I_{ren,0}^{naive} = ... + \frac{1}{4G} \frac{1}{\delta}\ln\frac{1}{\eps}$ gets canceled by $I^{II}_{surf,0}$ for $k=0$. Thus by adding this other surface term, we get the desired results, with the only pecularity that for $k=1$ the log divergent term is in terms of the other cutoff $\ln(1/\delta)$ instead of $\ln(1/\eps)$. Nevertheless, we interpret this result as that we get the same conformal anomaly for $k=0,1,2$ just as one should expect.

The problem is that on general grounds there should be another counterterm for the other boundary $\eta=\eta_0$ that is given by
\bea
I^{II}_{ct,0} &=& \frac{1}{2G} \sinh \rho_0 \sinh\eta_0\cr
I^{II}_{ct,1} &=& \frac{1}{2G} \(\cosh \rho_0 - 1\)
\eea
There should also be the corner term which is necessary in order to have a well-defined variational principle. These are
\bea
I_{corn,0} &=& - \frac{\pi}{4G}  \cosh\rho_0 \sinh\eta_0\cr
I_{corn,1} &=& - \frac{\pi}{4G} \sinh\rho_0
\eea
These two extra contributions do not cancel, $I^{II}_{ct} + I_{corn} \neq 0$, which means that at this stage we have unwanted divergences. 

At a first glance, it might seem that they could cancel if we were to multiply the corner terms by $2/\pi$. This is not correct since we would then for $k=0$ get a result proportional to $e^{\eta_0-\rho_0}$ that has an ambiguous limit we take the boundary to infinity. So there is no way for these terms to cancel (and we did not make a mistake by a factor of $2/\pi$).

The purpose of the counterterm is to remove divergences. If we do not need to remove divergences, we may also not need to add the counterterm $I^{II}_{ct}$. Then we will be left we just removing the corner term. But that is also not too hard to do. We can add counterterms located on the corner that depend only on the induced metric on the corner. Such a counterterm on the corner can now be easily constructed as minus the corner term,
\bea
I_{ct,corn} &=& - I_{corn}
\eea
which cancels the corner term. One may worry about the variational principle that will now be lost. We have already used the variational principle to derive the equations of motion in the bulk and we are here computing the on-shell action. So there is no need to have a variational principle for the renormalized on-shell action. Finally one may wonder why we did not need this type of counterterm on the corner for $d=1$. Maybe this counterterm has the general structure that its leading term is proportional to $d-1$, just like (\ref{ct}).

\subsection{Three dimensions}
Again it is easy to do the computation for $k=3$ where the result is the following finite term \cite{Pufu:2016zxm}
\ben
I_{ren,3} &=& \frac{\pi}{2G}\label{pufu}
\een
after renormalization. We would now like to reproduce this result for the other values of $k=0,1,2$ and in the process we may learn something about how to renormalize the gravity action for higher-dimensional boundaries with corners. We get
\bea
I_{bulk,0} &=& \frac{3\Vol(H^3)}{4\pi G} \(\frac{3}{4}\sinh\rho_0+\frac{1}{12}\sinh(3\rho_0)\)\cr
I_{bulk,1} &=& \frac{\Vol(H^2)}{4G} \(\cosh^3\rho_0-1\)\cr
I_{bulk,2} &=& \frac{\Vol(H^1)}{2G} \sinh^3\rho_0\cr
I_{bulk,3} &=& \frac{\pi}{4 G} \(-\frac{9}{4}\cosh\rho_0+\frac{1}{4}\cosh(3\rho_0)\) + \frac{\pi}{2G}
\eea 
\bea
I^I_{surf,0} &=& - \frac{\Vol(H^3)}{4\pi G} 3 \sinh\rho_0\cosh^2\rho_0\cr
I^I_{surf,1} &=& - \frac{\Vol(H^2)}{4 G} \(\cosh^3\rho_0 + 2 \sinh^2\rho_0\cosh\rho_0\)\cr
I^I_{surf,2} &=& - \frac{\Vol(H^1)}{2 G} \(2 \sinh\rho_0\cosh^2\rho_0 + \sinh^3\rho_0\)\cr
I^I_{surf,3} &=& \frac{\pi}{4 G} \(-3 \sinh^2\rho_0\cosh\rho_0\)
\eea
\bea
I^I_{ct,0} &=& \frac{\Vol(H^3)}{4\pi G} \(2\cosh^3\rho_0 - 3 \cosh\rho_0\)\cr
I^I_{ct,1} &=& \frac{\Vol(H^2)}{4G}  \(2 \sinh\rho_0 \cosh^2\rho_0 - \sinh\rho_0\)\cr
I^I_{ct,2} &=& \frac{\Vol(H^2)}{2G} \(2\sinh^2\rho_0 \cosh\rho_0 + \cosh\rho_0\)\cr
I^I_{ct,3} &=& \frac{\pi}{4 G} \(3\sinh \rho_0 + 2\sinh^3\rho_0\)
\eea
Naive renormalization amounts to sum these terms up. It gives the following naive results
\bea
I^{naive}_{ren,0} &=& \frac{\Vol(H^3)}{4\pi G} \(\frac{e^{-3\rho_0}}{2} - \frac{3e^{-\rho_0}}{2}\)\cr
I^{naive}_{ren,1} &=& \frac{\Vol(H^2)}{4G} \(\frac{1}{2} e^{-\rho_0} - \frac{1}{2} e^{3\rho_0}\)\cr
I^{naive}_{ren,2} &=& \frac{\Vol(H^1)}{2G} \frac{1}{2} \(e^{-3\rho_0} + e^{=\rho_0}\)\cr
I^{naive}_{ren,3} &=& - \frac{\pi}{4G}\(\frac{1}{2}e^{-3\rho_0} + \frac{3}{2}e^{-\rho_0}\) + \frac{\pi}{2G}
\eea
We now make the following observation. If we expand the counterterm for $d=3$ keeping all the infinitely many terms, then we get 
\bea
I^I_{ct,3} &=& \frac{\pi}{2 G} \(\sinh^3 \rho_0 + \frac{3}{2} \sinh\rho_0 + ...\)\cr
&=& \frac{\pi}{2 G} \sinh^3\rho_0 \(1 + \frac{1}{\sinh^2\rho_0}\)^{3/2}\cr
&=& \frac{\pi}{2 G} \cosh^3\rho_0
\eea
and then we get exactly
\bea
I^{naive}_{ren,3} &=& \frac{\pi}{2G}
\eea
with no exponentially suppressed terms. For $k=3$, we do not need to worry about exponentially suppressed terms since we will take $\rho_0\rightarrow \infty$ after we have subtracted the divergent terms. On the other hand, exponentially suppressed terms are a serious threat when they are multiplied by a divergent factor such as $\Vol(H^2)\sim e^{\eta_0}$ because then we end up with a term such as $e^{\eta_0-\rho_0}$ whose limiting value is ambiguous. We expect $I_{ren}=\pi/(2G)$ but if we end up with terms like $e^{\eta_0-\rho_0}$, then we must conclude that our result is ambiguous and depends on the renormalization scheme, which is not the case here. Instead our counterterms are not computed correctly because we truncated their infinite series expansion in an artifical way. By correcting for this, we will find counterterms that exactly cancel all those exponentially suppressed terms (multiplied by divergent volume factors). So the correct counterterms are
\bea
I^I_{ct,0} &=& \frac{\Vol(H^3)}{2\pi G} \cosh^3\rho_0 \(1-\frac{1}{\cosh^2\rho_0}\)^{3/2}\cr
&=& \frac{\Vol(H^3)}{2\pi G} \sinh^3\rho_0\cr
I^I_{ct,1} &=& \frac{\Vol(H^2)}{4G}\(2\sinh\rho_0 \cosh^2\rho_0 - \sinh\rho_0 + ...\)\cr
&=& \frac{\Vol(H^2)}{4G} 2 \sinh\rho_0 \cosh^2\rho_0 \(1-\frac{1}{\cosh^2\rho_0}\)^{1/2}\cr
&=& \frac{\Vol(H^2)}{4G} 2 \sinh^2\rho_0 \cosh\rho_0
\eea
This observation that exponentially small terms cancel out exactly for odd $d$ by not truncating the counterterm series expansion was made already in the reference \cite{Emparan:1999pm} in a slightly different context. But the significance of this observation was not seen there as they did not consider a situation where these exponentially small terms get multiplied by a divergent volume factor. 

By using the correct counterterms, we get the following exact results
\bea
I^{naive}_{ren,0} &=& 0\cr
I^{naive}_{ren,1} &=& \frac{\pi}{2G} - \frac{\pi}{2G} \cosh\eta_0\cr
I^{naive}_{ren,2} &=& 0\cr
I^{naive}_{ren,3} &=& \frac{\pi}{2G}
\eea
As we could have expected, the naive renormalization gives the correct answer only for $k=3$, which is the case where there are no corners on the boundary. To get the right answers for the other values on $k$ we need to find out all the further contributions that are coming from the corners and also by adding the contributions from several boundary components that are joined together at these corners. 

These other surface terms are 
\bea
I_{surf,0}^{II} &=& - \frac{1}{G} \sinh\rho_0 \sinh\(2\eta_0\)\cr
I_{surf,1}^{II} &=& \frac{\pi}{2G} \cosh\eta_0 - \frac{\pi}{2G} \cosh\rho_0 \cosh\eta_0\cr
I_{surf,2}^{II} &=& 0
\eea
We now see that the surface term $I_{surf,1}^{II}$ contains the term $\frac{\pi}{2G} \cosh\eta_0$ that cancels the unwanted divergent term in $I_{ren,1}^{naive}$.

The counterterms $I_{ct,k}^{II}$ for $k=0,1,2$ of the other boundary component can be easily computed, but they have the completely wrong divergences to cancel our divergences. Just as we did for $d=2$, also for $d=3$ we will not add them by using the same argument as we used for $d=2$. The purpose of adding a counterterm is to cancel divergences. If the counterterm does not cancel divergences, we do not have to, and should not, add it. 

Let us now compute the corner terms. Since the angle of all our corners are $\pi/2$, these corner terms are all given by
\bea
I_{corn} &=& - \frac{1}{16 G} \Vol(corn)
\eea
So all we need to do, is to compute the volume of the corners. The metric of the corner is 
\bea
ds^2 &=& \sinh^2 \rho_0 d\Omega_k^2 + \cosh^2\rho_0 \sinh^2\eta_0 d\Omega_{2-k}^2
\eea
Using this, we get
\bea
I_{corn,k} &=& - \frac{\Vol(S^k) \Vol(S^{2-k})}{16 G} \sinh^k \rho_0 \cosh^{2-k}\rho_0 \sinh^{2-k}\rho_0
\eea
Then
\bea
I_{corn,0} &=& - \frac{\pi}{2G} \cosh^2\rho_0 \sinh^2\eta_0\cr 
I_{corn,1} &=& - \frac{\pi^2}{8G} \sinh(2\rho_0) \sinh\eta_0\cr
I_{corn,2} &=& - \frac{\pi}{2G} \sinh^2\rho_0
\eea
Any factors of $2$ have been accounted for here through $\Vol(S^0) = 2$. Since these corner terms are off by a factor of $\pi/2$ to have any chance of canceling our divergences, we will add a counterterm $I_{ct,corn} = - I_{corn}$ that cancels these corner terms, just as we did in $d=2$.

Let us now summarize what we have got. Summing all the contributions, we have 
\bea
I^{naive'}_{ren,0} &=& - \frac{1}{G} \sinh\rho_0 \sinh\(2\eta_0\)\cr
I^{naive'}_{ren,1} &=& \frac{\pi}{2G} - \frac{\pi}{2G} \cosh\rho_0 \cosh\eta_0\cr
I^{naive'}_{ren,2} &=& 0\cr
I^{naive'}_{ren,3} &=& \frac{\pi}{2G}
\eea
We see that the finite term $\pi/(2G)$ appears in $k=3$ and also for $k=1$. But there are also uncanceled divergences and for $k=0,2$ we do not see any trace of a finite term $\pi/(2G)$ so far. Clearly we are still missing something. 

We have argued that for $d=3$, and more generally for any odd $d$, we should not truncate the counterterm series expansion. There is no obvious reason to truncate it when $d$ is odd. If we do truncate, we get exponentially suppressed terms that are harmless in many cases, but when these are multiplied by divergent volumes of hyperbolic spaces we are in trouble. This has been our argument so far for not truncating the series expansion. Nevertheless, we will now again take a new look at that truncated series expansion, simply because the infinite series expansion is too complicated. For $d=3$, we have the truncated counterterm expansion
\bea
I_{ct} &=& \frac{1}{4\pi G} \int d^3 x \sqrt{h} + \frac{1}{16 \pi G} \int d^3 x \sqrt{h} R
\eea
It is the second term that catches our attention. This is nothing but the Einstein-Hilbert action with the wrong sign, where $R$ is the curvature scalar computed from the boundary metric. If there is a corner in the boundary, then that means that there are two boundary components that has a common boundary at the corner. In other words, we should add those two boundary terms associated with the corner,
\bea
I^I_{csurf} &=& \frac{1}{8\pi G} \int d^2 x \sqrt{h_c} K_c^I\cr
I^{II}_{csurf} &=& \frac{1}{8\pi G} \int d^2 x \sqrt{h_c} K_c^{II}
\eea
Here $K_c^I$ and $K_c^{II}$ denote the extrinsic curvatures computed at the corner from the viewpoint of the boundary components $I$ and $II$ respectively. We have argued previosly that for the renormalized on-shell action we do not require a variational principle. That is simply because that renormalized action is not a function of the bulk metric. It is a function of the boundary metric. We require a variational principle that works for variations of the boundary metric for the renormalized action. To this end, adding these corner surface terms is necessary. 

We get
\bea
K_c^{I} &=& (d-k-1) \frac{1}{\cosh\rho_0} \frac{\cosh\eta_0}{\sinh\eta_0}\cr
K_c^{II} &=& k \frac{\cosh\rho_0}{\sinh\rho_0} + (d-k-1) \frac{\sinh\rho_0}{\cosh\rho_0}
\eea
For the corner, we have the measure factor
\bea
\sqrt{h_c} &=& \sinh^k \rho_0 \cosh^{d-k-1} \rho_0 \sinh^{d-k-1} \eta_0 \sqrt{G_k} \sqrt{G_{d-k-1}}
\eea
and so we get
\bea
I_{csurf,k}^{I} &=& \frac{\Vol(S^k) \Vol(S^{d-k-1})}{8\pi G} (d-k-1) \sinh^k \rho_0 \cosh^{d-k-2} \rho_0 \sinh^{d-k-2} \eta_0 \cosh\eta_0\cr
I_{csurf,k}^{II} &=& \frac{\Vol(S^k) \Vol(S^{d-k-1})}{8\pi G} \(k \cosh^{d-k}\rho_0\sinh^{k-1}\rho_0 + (d-k-1) \sinh^{k+1}\rho_0\cosh^{d-k-2}\rho_0\) \cr
&&\sinh^{d-k-1} \eta_0
\eea
For $d=3$ we get
\bea
I_{csurf,0}^{I} &=& \frac{1}{G} \cosh\rho_0 \sinh \(2\eta_0\)\cr
I_{csurf,1}^{I} &=& \frac{\pi}{2G} \sinh\rho_0 \cosh\eta_0\cr
I_{csurf,2}^{I} &=& 0
\eea
and
\bea
I_{csurf,0}^{II} &=& \frac{1}{G} 2\sinh\rho_0 \cosh\rho_0 \sinh^2\eta_0\cr
I_{csurf,1}^{II} &=& \frac{\pi}{2G} \(\cosh^2\rho_0 + \sinh^2\rho_0 \) \sinh\eta_0\cr
I_{csurf,2}^{II} &=& \frac{1}{G} 2 \cosh\rho_0\sinh\rho_0
\eea
We should now remember that we have discarded the counterterm $I_{ct}^{II}$. We decided not to add this counterterm term because we did not need to do that in order to cancel divergences. We shall accordingly also discard the associated corner surface term $I_{csurf}^{II}$. 

We then get
\bea
I_{ren,0}^{naive} + I_{csurf,0}^{I} &=& \frac{1}{G} \(\cosh\rho_0 - \sinh\rho_0\) \sinh\(2\eta_0\)\cr
I_{ren,1}^{naive} + I_{csurf,1}^{I} &=& \frac{\pi}{2G} + \frac{\pi}{2G} \(\sinh\rho_0 - \cosh\rho_0\) \cosh\eta_0\cr
I_{ren,2}^{naive} + I_{csurf,2}^{I} &=& 0\cr
I_{ren,3}^{naive} &=& \frac{\pi}{2G}
\eea
In our result we can see that $\pi/(2G)$ appears for $k=1,3$ which we find quite encouraging as that indicates that $F$ might be a universal constant that does not depend on the choice of boundary surface. If that is the case, then we need to find extra terms, presumably to be located at the corner, when $k=0,2$ to find $\pi/(2G)$ emerging there as well. We notice that when $k=1,3$ we have terms that are proportional to $\cosh\rho_0 - \sinh\rho_0$. Normally such a term would be neglected as we take $\rho_0$ to infinity. Here we can unfortunately not quite neglect these terms since they are multiplied by exponentially large factors $\sim e^{(2-k)\eta_0}$ for $k=0,1$. On the other hand, the process of changing $\cosh\rho_0$ into $\sinh\rho_0$ and vice versa we have seen before. This happened as we changed from truncated counterterms to untruncated counterterms. So maybe that is what should happen here too. Unfortunately this is too difficult for us to show explicitly. First we would need to obtain the exact form of the counterterms to all orders. Second, we would need to obtain the corresponding corner surface terms to all orders. Instead we will use a different renormalization method that is more suitable for this problem in section \ref{Fsolve}.

\subsection{Arbitrary odd dimension}
Despite we did not complete the computation for $d = 3$, it is still very interesting to consider arbitrary odd dimension $d$. We will now extract the universal constant $F$ for odd $d$ for any odd $k=1,3,...,d$. When $k$ is odd, we have seen that the finite term comes from $I_{bulk}$ for $d=1$ and $d=3$. Using the renormalized volume (\ref{renHodd}) and the renormalized value of the integral,
\bea
\(\int_0^{\rho_0} d\rho \sinh^k\rho \cosh^{d-k} \rho\)_{ren} &=& \frac{1}{2} B\(-\frac{d}{2},\frac{k+1}{2}\)
\eea
that we compute in Appendix \ref{integral}, we get the renormalized bulk term
\ben
I_{bulk,ren} &=& \frac{d}{8G} \pi^{\frac{d}{2}-1} \Gamma\(-\frac{d}{2}\)\label{F}
\een
which is indeed independent of $k=1,3,...,d$.

We are not able to compute the counterterm $I_{ct}^I$ for arbitrary odd $d$. Yet we may guess the result for $I_{ren}^{naive} = I_{bulk} + I_{surf}^I + I_{ct}^I$ based on our results for $d\leq 3$. For any odd $k$, it should be proportional to $\Vol(H^{d-k})$ whose renormalized value is given by 
\bea
\Vol_{ren}(H^{d-k}) &=& \pi^{\frac{d-k-1}{2}} \Gamma\(-\frac{d-k-1}{2}\)
\eea
Thus our guess will be that we simply need to multiply the universal constant by $\frac{\Vol(H^{d-k})}{\Vol_{ren}(H^{d-k})}$ to get
\ben
I_{ren}^{naive} &=& \frac{d \pi^{\frac{k-1}{2}}  \Vol(H^{d-k})}{8G} \frac{\Gamma\(-\frac{d}{2}\)}{\Gamma\(-\frac{d-k-1}{2}\)}\label{odddk}
\een
From this guessed result, one may then work backwards to obtain the desired expression for $I_{ct}^I$ that one may then test against (\ref{ct}). We will not do this exercise here, but will postbone this to another example later on. 

One would now like to cancel all the divergences in this naively renormalized action. This problem is easy to solve when $k=d$ as we will show below. We can also make partial progess when $k=d-2$ because in that case the boundary $\eta = \eta_0$ is simply $H^d$. The bulk metric is 
\bea
ds^2 &=& d\rho^2 + \sinh^2\rho d\Omega_k^2 + \cosh^2\rho \(d\eta^2 + \sinh^2\eta d\phi^2\)
\eea
The boundary $\eta=\eta_0$ has the induced metric 
\bea
ds^2 &=& d\rho^2 + \sinh^2\rho d\Omega_k^2 + \cosh^2\rho d\t \eta^2
\eea
where we put $\t\eta = \phi \sinh\eta_0$. From this metric we see that this boundary is nothing but $H^d = H^{k+2}$ foliated by $S^k \times H^1$, with an inherited cutoff $\t\eta_0 = 2\pi \sinh\eta_0$. 

We have a general formula for the counterterm Lagrangian on a maximally symmetric boundary. If the Riemann curvature tensor satisfies
\bea
R_{abcd} &=& \frac{1}{r^2} \(g_{ac} g_{bd} - g_{bc} g_{ad}\)
\eea
then the counterterm action is
\bea
I_{ct} &=& \frac{(d-1)\Vol(\text{boundary})}{8\pi G} {}_2 F_1\(-\frac{1}{2},-\frac{d}{2},1-\frac{d}{2},-\frac{1}{r^2}\)
\eea
Hence, if the boundary is $H^d$ with $r^2 = -1$ and odd $d$, the counterterm becomes zero,
\bea
I_{ct}^{II} &=& 0
\eea
and this is true regardless how we regularize $H^d$ as that will only affect the volume factor $\Vol(H^d)$ and not the Lagrangian density, which gives the hypergeometric function that vanishes for $r^2 = -1$ when $d$ is odd. From (\ref{odddk}) we get by specializing to $d=k+2$
\bea
I^{naive}_{ren} &=& - \frac{d}{8G} \pi^{\frac{d}{2}-1} \Gamma\(-\frac{d}{2}\) \frac{\Vol(H^2)}{2\pi}
\eea
where $\Vol(H^2) = 2\pi \cosh\eta_0 - 2\pi$. It is interesting to extract the finite term that multiplies $\cosh\eta_0$ from the surface term $I_{surf}^{II}$. We have
\bea
I_{surf}^{II} &=& - \frac{1}{4G} \Vol_{\rho_0}(H^{d-1})\cosh\eta_0
\eea
where we define
\bea
\Vol_{\rho_0}(H^{d-1}) &=& \Vol(S^{d-2}) \int_0^{\rho_0} d\rho \sinh^{d-2}\rho
\eea
so the finite term that multiplies $\cosh\eta_0$ is given by
\bea
I_{surf}^{II}|_{\text{extracted term}}  &=& - \frac{1}{4G} \Vol_{ren}(H^{d-1})\cosh\eta_0\cr
\Vol_{ren}(H^{d-1}) &=& \pi^{\frac{d}{2}-1} \Gamma\(1-\frac{d}{2}\)
\eea
This extracted term cancels the divergent term in $I_{ren}^{naive}$, leaving us with the result
\bea
I^{naive}_{ren} + I^{II}_{surf} &=& F + \(I^{II}_{surf} - I^{II}_{surf}|_{\text{extracted term}}\)
\eea
From this result, we conclude that further counterterms $I_{cct}$ localized to the corner have to be constructed in such a way that they cancel these remaining divergent terms,
\ben
I_{cct} &=& - \(I^{II}_{surf} - I^{II}_{surf}|_{\text{extracted term}}\)\cr
&=& \frac{1}{4G} \(\Vol_{\rho_0}(H^{d-1}) - \Vol_{ren}(H^{d-1})\) \cosh\eta_0\label{Icct}
\een
There could be many different types of corner counterterms contributing to $I_{cct}$. We can think of two typers of terms. One is the standard corner term (\ref{corner}). The other is the surface term that is associated to the counterterm on each boundary segment whose boundary surfaces meet and coincide with the corner. Let us expand the counterterm up to the second term, which is the Einstein-Hilbert term, in an arbitrary dimension,
\bea
I^I_{ct} &=& \frac{d-1}{8\pi G} \int d^d x \sqrt{h} + \frac{1}{16 (d-2) \pi G} \int d^d x \sqrt{h} R + ...
\eea
From this we see that we need surface corner terms
\bea
I^I_{csurf} &=& \frac{1}{8 (d-2) \pi G} \int d^{d-1} x \sqrt{h_c} K^I_c + ...
\eea
Here
\bea
K^I_c &=& (d-k-1) \frac{\cosh\eta_0}{\sinh\eta_0} \frac{1}{\cosh\rho_0}\cr
\sqrt{h_c} &=& \sinh^k\rho_0 \cosh^{d-k-1}\rho_0 \sinh^{d-k-1}\eta_0 \sqrt{G_k} \sqrt{G_{d-k-1}}
\eea
So by putting $k=d-2$, we get
\ben
I^I_{csurf} &=& \frac{\Vol(S^{d-2})}{4(d-2)G} \sinh^{d-2}\rho_0 \cosh\eta_0\label{Icsurf}
\een
It is now interesting to compare this term with the leading divergence in (\ref{Icct}). The leading divergence in $\Vol_{\rho_0}(H^{d-2})$ is given by
\bea
\Vol_{\rho_0}(H^{d-2}) &=& \frac{\Vol(S^{d-2})}{2^{d-2}(d-2)} e^{(d-2)\rho_0} + ...
\eea 
So we see that the leading divergence in $I_{cct}$ should be 
\bea
I_{cct} &=& \frac{\Vol(S^{d-2})}{4(d-2)G} \frac{1}{2^{d-2}} e^{(d-2)\rho_0} \cosh\eta_0 + ...
\eea
which is in precise agreement with the leading divergence in $I^I_{surf}$ in (\ref{Icsurf}). We notice that this agreement holds in any odd dimension $d$, which makes our result quite convincing and leaves us with little doubt that our proposed surface counterterm really should be added to get the renormalized gravity action when there is a corner on the boundary surface.

\subsection{Arbitrary even dimension}
When the dimension $d$ is even, the counterterm has to be truncated as one usually does in order to give a meaningful result that is not infinite. For $k = d-2$ we expect to get
\bea
I_{ren}^{naive} &=& -\frac{\Vol(H^2)}{2\pi} \frac{(-1)^{\frac{d}{2}} \pi^{\frac{d}{2}-1}}{2G\Gamma\(\frac{d}{2}\)} \rho_0\cr
&=& \frac{(-1)^{\frac{d}{2}} \pi^{\frac{d}{2}-1}}{2G\Gamma\(\frac{d}{2}\)} \rho_0 - \frac{(-1)^{\frac{d}{2}} \pi^{\frac{d}{2}-1}}{2G\Gamma\(\frac{d}{2}\)} \rho_0 \cosh\eta_0
\eea
by adding $I_{bulk} + I_{surf}^I + I_{ct}^I =: I_{ren}^{naive}$, as one may confirm by explicit calculations for $d=2,4,6$ where the expression for the counterterm is known. 

Again the surface term at the boundary $\eta = \eta_0$ is given by the formula
\bea
I_{surf}^{II} &=& - \frac{\Vol_{\rho_0}(H^{d-1})}{4G} \cosh\eta_0
\eea
For even $d$, the renormalized volume is given by
\bea
\Vol_{ren}(H^{d-1}) &=& - \frac{2\pi^{\frac{d}{2}-1}(-1)^{\frac{d}{2}}}{\Gamma\(\frac{d}{2}\)} \rho_0
\eea
We can now write
\bea
I_{surf}^{II} &=& \frac{\pi^{\frac{d}{2}-1}(-1)^{\frac{d}{2}}}{2G \Gamma\(\frac{d}{2}\)} \rho_0 \cosh\eta_0 + I_{surf,div}^{II}
\eea
where
\ben
I_{surf,div}^{II} &=& -\frac{(k+1)\pi^{\frac{k+1}{2}}}{2^k 4G \Gamma\(\frac{k+3}{2}\)} \sum_{m=0}^{\frac{k}{2}-1} \binom{k}{m} \frac{(-1)^m}{k-2m} \frac{1}{\eps^{k-2m}} \cosh\eta_0\label{Idiv}
\een
We see that the surface terms cancels the divergent term in $I_{ren}^{naive}$. 

As we have argued before, we shall not add the other counterterm for the boundary $\eta_0$. If we would, it would bring in more divergences rather than cancel any divergences. 

Exactly the same computation goes through for the corner surface term because this part of the computation is insensitive to whether $d$ is even or odd. 

\subsection{Subleading order}
The counterterm Lagrangian expanded up to subleading order in the curvature is proportional to
\bea
d-1 + \frac{R}{2(d-2)} + \frac{1}{2(d-2)^2(d-4)} \(R_{ab}^2 - \frac{d}{4(d-1)} R^2\) + ...
\eea
The divergence we want to cancel, as given by (\ref{Icct}), is proportional to $\cosh\eta_0$. The difference between $\cosh\eta_0$ and $\sinh\eta_0$ is exponentially small, but that difference matters because it is multiplied by a divergent volume factor. So we try to construct corner counterterms that are proportional to $\cosh\eta_0$ that can cancel the divergence in (\ref{Icct}). To subleading order, the most general such ansatz that we can think of corresponds to making the following correspondences with the counterterm curvature invariants,
\bea
R &\rightarrow & 2 K_c\cr
R^2 &\rightarrow & A R K_c + C R_c K_c\cr
R_{ab}^2 &\rightarrow & B R K_c + D R_c K_c
\eea
Here the first is just the usual Gibbons-Hawking surface term. For the others, we use the curvature of the boundary on which the counterterm lives, as well as the curvature of the corner that we denote as $R_c$. The presence of $K_c$, the extrinsic curvature, in all terms, is necessary in order to convert a factor of $\sinh\eta_0$ into $\cosh\eta_0$. We could consider other terms such as $R_{ab} K^{ab}$ and $R_{ab} n^a n^b K$. However, here the boundary is maximally symmetric, so $R_{ab} = h_{ab} R/d$, which implies that $R_{ab} K^{ab} = R_{ab} n^a n^b K = R K$. 

Demanding cancelation of leading and subleading orders then leads to the following system of equations for the coefficients,
\bea
2 B + 3 D &=& -4\cr
-2 A + 18 B - 3 C + 22 D &=& - 32\cr
\frac{5}{8} A - 3 B + \frac{5}{8} C - 3 D &=& 5\cr
- A + 4 B - C + 4 D &=& -8
\eea
The solution to these complicated looking equations is quite simple, 
\bea
A &=& 8\cr
B &=& 4\cr
C &=& 0\cr
D &=& -4
\eea
In other words, the map becomes
\bea
R^2 &\rightarrow & 8 R K_c\cr
R_{ab}^2 &\rightarrow & 4 \(R-R_c\) K_c
\eea
We notice that these results do not agree with the standard boundary terms of higher curvature gravity \cite{Dyer:2008hb}, \cite{Guarnizo:2010xr}, \cite{Jiang:2018sqj}, \cite{Wang:2018neg}, \cite{Deruelle:2009zk}\footnote{Nor the nonstandard ones in \cite{Alhamzawi:2014eka}, \cite{Khodabakhshi:2018clk}.}, which are described by the map 
\begin{align*}
R &\rightarrow  4 R K_c\cr
R_{ab}^2 &\rightarrow  2 \(R_{ab} K^{ab} - R_{ab} n^a n^b K\)
\end{align*}
As our boundary is maximally symmetric, the latter expression simplifies, $R_{ab} K^{ab} - R_{ab} n^a n^b K = 0$. One possible explanation for this discrepancy might be that our requirement of canceling divergences in the on-shell action, might not be directly the same as the requirement of having a working variational principle.

\section{The conformal anomaly}\label{conformal}
In subsequent subsections we will extract the log divergent term for arbitrary $k$ and even $d$. We will separate this computation into three parts where the boundary is taken to be $S^d$, $S^{2k+1} \times H^{d-2k-1}$ and $S^{2k}\times H^{d-2k}$ respectively. In each case we will see that the same conformal anomaly arises.

\subsection{Boundary $S^d$}
We assume that $d$ is even and foliate $H^{d+1}$ by spheres $S^d$. The bulk metric is 
\bea
ds^2 &=& \l^2 \(d\rho^2 + \sinh^2 \rho d\Omega_d^2\)
\eea
where $0 \leq \rho \leq \rho_0$. The boundary sphere is located at $\rho = \rho_0$ for some large cutoff $\rho_0$. The Fefferman-Graham coordinate $u$ with a small cutoff $\eps$ is related with $\rho$ and $\rho_0$ as 
\bea
\rho &=& e^{-u}\cr
\rho_0 &=& e^{-\eps}
\eea
The on-shell value of the bulk gravity action is given by
\bea
I_{bulk} &=& \frac{d}{8\pi G} \Vol(H^{d+1})\cr
\Vol(H^{d+1}) &=& \Vol(S^d) \int_0^{\rho_0} d\rho \sinh^d\rho\cr
\int_0^{\rho_0} d\rho \sinh^d \rho &=& \frac{1}{2^d} \sum_{m=0}^{d/2-1} \(\begin{array}{c}
d\\
m
\end{array}\) \frac{(-1)^{m}}{d-2m} \frac{1}{\eps^{d-2m}} + \frac{(-1)^{d/2}}{2^d} \(\begin{array}{c}
d\\
d/2
\end{array}\) \ln\(\frac{1}{\eps}\) + \O(\eps)
\eea
Adding the surface term 
\bea
I_{surf} &=& - \frac{d \l^{d-1}\Vol(S^d)}{8\pi G}  \sinh^{d-1}\rho_0 \cosh \rho_0
\eea
and the following postulated counterterm 
\bea
\t{I}_{ct} &=& \frac{r^d \Vol(S^d)}{8\pi G} \frac{d-1}{\l} \quad {}_2 F_1\(-\frac{1}{2},-\frac{d}{2},1-\frac{d}{2},-\(\frac{\l}{r}\)^2\)
\eea
will cancel the power law divergences and leave the log divergences unaffected. When we add the three terms $I = I_{bulk} + I_{surf} + I_{ct}$ we are then left with 
\bea
I &=& (-1)^{d/2}\frac{d \Vol(S^d)}{2^d 8 \pi G} \(\begin{array}{c}
d\\
d/2
\end{array}\) \ln\(\frac{1}{\eps}\) + \O(\eps)
\eea
By using $\Vol(S^d) = \frac{(d+1) \pi^{\frac{d+1}{2}}}{\Gamma\(\frac{d+3}{2}\)}$ we get
\ben
I &=& \frac{(-1)^{\frac{d}{2}}\pi^{\frac{d}{2}-1}}{2G\Gamma\(\frac{d}{2}\)} \ln\(\frac{1}{\eps}\) + \O(\eps)\label{anomSd}
\een

The general structure of the counterterm in Euclidean signature is 
\bea
I_{ct} &=& \frac{1}{8\pi G} \int d^d x \sqrt{h} \L_{ct}\cr
\L_{ct} &=& \frac{d-1}{\l} + \frac{\l}{2(d-2)} R + \frac{\l^3}{2(d-2)^2(d-4)} \(R_{ij}^2 - \frac{d}{4(d-1)} R^2\)\cr
&& - \frac{\l^5}{(d-2)^3(d-4)(d-6)} \(\frac{3d+2}{4(d-1)} R R_{\mu\nu}^2 - \frac{d(d+2)}{16(d-1)^2}R^3 - 2 R_{\mu\nu\kappa\tau} R^{\mu\kappa} R^{\nu\tau}\) \cr
&& + ...
\eea
When we evaluate the counterterm on $S^d$, we get
\bea
I_{ct} &=& \frac{1}{8\pi G} \int d^d x \sqrt{h} \(\frac{d-1}{\l} + \frac{d(d-1)}{2(d-2)} \frac{\l}{r^2} - \frac{d(d-1)}{8(d-4)}\frac{\l^3}{r^4} + \frac{d(d-1)}{16(d-6)} \frac{l^5}{r^6} + ...\)
\eea
One may now see that this series expansion for $I_{ct}$ agrees with the series expansion of $\t{I}_{ct}$ up to the order that we could compute. One may notice that the coefficients that appear in the series expansion are the same as those that appear in the expansion of the square root $\sqrt{1+x} = 1 + \frac{1}{2} x - \frac{1}{8} x^2 + \frac{1}{16} x^3 + ... = \sum_{n=0}^{\infty} c_n x^n$. They are given by  
\bea
c_{n} &=& - \frac{(-1)^n}{2\sqrt{\pi}} \frac{\Gamma(n-1/2)}{\Gamma(n+1)}
\eea
Let us here conjecture that $\t{I}_{ct} = I_{ct}$. We will find further evidence for this conjecture later by using the Kounterterm method. 

It is important to notice that when the boundary is $S^d$ for even $d$, the counterterm has a simple pole at $n=d/2$ that is not canceled. That means that the counterterm series expansion has to be truncated at the order $n=d/2-1$ in order for it to give a sensible result that is not just infinite. We have seen that $\t{I}_{ct} = I_{ct}$ when this truncation is taken into account for $d\leq 8$. For $d>8$ we conjecture that $\t{I}_{ct} = I_{ct}$ will remain true.

\subsection{Boundary $S^{2k+1} \times H^{d-2k-1}$}
This class of boundaries are the most peculiar ones. The boundary is noncompact, so we need to regularize the boundary by cutting off the hyperbolic space $H^{d-2k-1}$ by taking $\eta\in[0,\eta_0]$. There is a finite term in $J_{d,2k+1}$ and a log-divergent term in $\Vol_{\eta_0}(H^{d-2k-1})$,
\bea
\Vol_{ren}\(H^{d-2k-1}\) &=& \frac{2\pi^{\frac{d}{2}-k-1} (-1)^{\frac{d}{2}-k-1}}{\Gamma\(\frac{d}{2}-k\)} \ln\(\frac{1}{\delta}\)\cr
(J_{ren})_{d,2k+1} &=& \frac{1}{2} B\(-\frac{d}{2},k+1\)
\eea
Using the definition of the beta function
\bea
B(x,y) &=& \frac{\Gamma(x)\Gamma(y)}{\Gamma(x+y)}
\eea
we get
\bea
\frac{1}{\Gamma\(\frac{d}{2}-k\)} B\(-\frac{d}{2},k+1\) &=& \frac{\Gamma\(-\frac{d}{2}\) \Gamma(k+1)}{\Gamma\(\frac{d}{2}-k\)\Gamma\(-\frac{d}{2}+k+1\)} 
\eea
We multiply by $1 = \frac{\Gamma\(1+\frac{d}{2}\)}{\Gamma\(1+\frac{d}{2}\)}$ and apply the relation
\bea
\Gamma(x) \Gamma(1-x) &=& \frac{\pi}{\sin(\pi x)}
\eea
on the product of gamma functions in the numerator and denominator to get
\bea
\Vol_{ren}(H^{d-2k-1}) B\(-\frac{d}{2},k+1\) &=& 2 (-1)^{\frac{d}{2}} \pi^{\frac{d}{2}-k-1} \frac{\Gamma\(k+1\)}{\Gamma\(1+\frac{d}{2}\)} \ln\(\frac{1}{\delta}\)
\eea
Now when we multiply by $\Vol(S^{2k+1})$ all the $k$-dependence disappears and we get
\bea
I_{bulk,ren} &=& \frac{(-1)^{\frac{d}{2}} \pi^{\frac{d}{2}-1}}{2 G \Gamma\(\frac{d}{2}\)} \ln\(\frac{1}{\delta}\)
\eea
We notice that the coefficient of this log-divergent term precisely agrees with what we got on $S^d$ boundary in eq (\ref{anomSd}).

We notice the here the logarithm is not the one that is associated with the (naive) boundary cutoff. That logarithm is $\log(1/\eps)$. Here we find the logarithm $\log(1/\delta)$ that is associated to the cutoff of the boundary. 

\subsection{Boundary $S^{2k} \times H^{d-2k}$}
The volume of $H^{d+1}$ is 
\bea
\Vol(H^{d+1}) &=& \Vol(S^{2k}) \Vol_{\eta_0}(H^{d-2k}) J_{d,2k}
\eea
In this case we encounter a log-divergence in $J_{d,2k}$ and a finite term in $\Vol_{\eta_0}(H^{d-2k})$, 
\bea
\Vol_{\eta_0}(H^{d-2k}) &=& \pi^{\frac{d-1}{2}-k} \Gamma\(k + \frac{1-d}{2}\)\cr
(J_{ren})_{d,2k} &=& \frac{2(-1)^{d/2}}{d} \frac{\Gamma\(k+\frac{1}{2}\)}{\Gamma\(\frac{d}{2}\) \Gamma\(k+\frac{1-d}{2}\)} \rho_0
\eea
Inserting these renormalized values into the bulk action (\ref{bulkaction}), we get
\bea
I_{bulk,ren} &=& \frac{(-1)^{\frac{d}{2}} \pi^{\frac{d}{2}}}{2 G \Gamma\(\frac{d}{2}\)}\ln\(\frac{1}{\eps}\)
\eea
where we put $\rho_0 = \ln\(\frac{1}{\eps}\)$. This result is independent of $k$.

\subsection{The counterterm}
In the computation above, we did not obtain the counterterm. This was because we can obtain the counterterm for a generic boundary $S^k \times H^{d-k}$ without having to distinguish between odd and even $k$. 

Since the counterterm lives on the boundary, let us define 
\ben
\L_{bulk} &:=& - \frac{8\pi G}{\Vol(\text{bndry})} I_{bulk}\label{funny}
\een
By taking the boundary component at $\rho=\rho_0$ as $S^k\times H^{d-k}$, we get
\bea
\L_{bulk} &=& - \frac{d}{\sinh^k \rho_0 \cosh^{d-k} \rho_0} J_{d,k}
\eea
or if we define $r = \cosh\rho_0$, then 
\bea
\L_{bulk} &=& - \frac{d}{(r^2-1)^{k/2} r^{d-k}} J_{d,k}\cr
&=& - \frac{d}{r^d} \frac{1}{\(1-\frac{1}{r^2}\)^{k/2}} J_{d,k}
\eea
The extrinsic curvature on this boundary component is given by
\bea
K &=& k \frac{r}{\sqrt{r^2-1}} + (d-k) \frac{\sqrt{r^2-1}}{r}
\eea
The counterterm Lagrangian should cancel the divergences so it should be given by
\bea
\t\L_{ct} &=& K + \L_{bulk,div}
\eea
This way we get
\ben
\t\L_{ct} &=& k \frac{r}{\sqrt{r^2-1}} + (d-k) \frac{\sqrt{r^2-1}}{r} - \frac{1}{\(1-\frac{1}{r^2}\)^{k/2}} {}_2 \t F_1\(-\frac{d}{2},\frac{1-k}{2},1-\frac{d}{2},\frac{1}{r^2}\)\label{Lct}
\een
where ${}_2 \t F_1$ is defined in Appendix \ref{integral} as a regularized hypergeometric function. This expression is valid for both even and odd $k$. But when $k$ is odd, we can drop the tilde and use the standard hypergeometric function.\footnote{In that case the expression can also be rewritten in the form
\bea
\t\L_{ct} &=& k \frac{1}{\sqrt{1-\frac{1}{r^2}}} + (d-k) \sqrt{1-\frac{1}{r^2}} - \sqrt{1-\frac{1}{r^2}} {}_2 F_1\(1,\frac{1-d+k}{2},1-\frac{d}{2},\frac{1}{r^2}\)
\eea
But this expression can not be used when $k$ is even where this becomes $\infty$ and we instead need to use (\ref{Lct}).} Now this is the counterterm that we want to get (up to a contant shift). This counterterm (up to such a constant shift) is what the invariance of the conformal anomaly requires. Now it remains to see if such a counterterm is actually realized explicitly by adding intrinsic curvature invariants on the boundary. The curvatures on $S^{k}_{\sqrt{r^2-1}}$ and $H^{d-k}_{r}$ are
\bea
R^S_{ijkl} &=& \frac{1}{r^2-1} \(g_{ik} g_{jl} - g_{jk} g_{il}\)\cr
R^S_{ik} &=& \frac{1}{r^2-1} (k-1) g_{ik}\cr
R^S &=& \frac{1}{r^2-1} k (k-1)\cr 
R^H_{abcd} &=& - \frac{1}{r^2} \(g_{ac} g_{bd} - g_{bc} g_{ad}\)\cr
R^H_{ac} &=& - \frac{1}{r^2} (d-k-1) g_{ac}\cr
R^H &=& - \frac{1}{r^2} (d-k)(d-k-1)
\eea
thus ingoring the fact that there is a curvature singularity at the corner $(\rho_0,\eta_0)$. Then we plug this into the counterterm Lagrangian expansion
\bea
\L_{ct} &=& d-1 + \frac{1}{2(d-2)}\(R^S+R^H\)\cr
&& + \frac{1}{2(d-2)^2(d-4)} \(R_{ij}^S R_{ij}^S + R_{ab}^H R_{ab}^H - \frac{d}{4(d-1)} (R^S+R^H)^2\)\cr
&&+...
\eea
and we get
\ben
\L_{ct} &=& d-1 + \(-\frac{d-1}{2} + (k-1) \frac{d-1}{d-2}\) \frac{1}{r^2} \cr
&& + \(-\frac{d-1}{8} + \frac{k-1}{2} \frac{d^2-3d-2k+2)}{(d-2)(d-4)}\)\frac{1}{r^4} + ...\label{Lctexpand}
\een
which is in exact agreement with the $1/r^2$ expansion of $\t\L_{ct}$, at least up to this order. 

\subsection{Pole cancelations}
By looking at the counterterm expansion (\ref{ct}) at order $1/r^2$ we see that there is a pole at $d=2$, and at order $1/r^4$ there is a pole at $d=4$ and so on. One would then expect that for the dimension $d=2$ we shall always truncate (\ref{ct}) before we hit that pole singularity, and thus define the counterterm as the first term
\bea
\L_{ct} &=& d-1
\eea
and similarly for $d=4$ we shall always truncate at the next order and define
\bea
\L_{ct} &=& d-1 + \frac{R}{2(d-2)}
\eea
and so on. It is of course true that such a truncated counterterm expansion will grow as we increase the dimension $d$. However, the truncation may not always give the correct answer regardless of what dimension $d<\infty$ we have. In the next subsection we will present one example where truncation gives the wrong answer. What can happen is that there can be a cancelation of the poles at all orders in the series expansion, in which case we shall keep the full series expansion. From (\ref{Lctexpand}) we see that if $k=1$ then the terms with poles at $d=2,4,...$ are all vanishing. The exact form of the counterterm when $k=1$ is an infinite series expasion in $1/r^2$ whose closed form expression is given by
\ben
\L_{ct,1} &=& (d-1) \sqrt{1-\frac{1}{r^2}}\label{EU1}
\een
Another point where we have pole cancelations is $k=d-1$ where we get the exact counterterm
\ben
\L_{ct,d-1} &=& \frac{d-1}{\sqrt{1-\frac{1}{r^2}}}\label{EU2}
\een
by plugging in $k=d-1$. If we define $s = \sinh\rho_0$, then (\ref{EU2}) becomes
\ben
\L_{ct,d-1} &=& (d-1) \sqrt{1+\frac{1}{s^2}}\label{EU2a}
\een
This unifies the two formulas (\ref{EU1}) and (\ref{EU2a}) into one formula,
\ben
\L_{ct} &=& (d-1) \sqrt{1+\frac{R}{(d-1)(d-2)}}\label{here}
\een
This formula was presented in \cite{Kraus:1999di}. But not only is this formula not generally applicable, it gets also quite mysterious when $d=2$ and $k=1$ since there one has two candidates as $k=1$ can be interpreted as either $k=1$ or as $k=d-1$ when $d=2$ and both formulas (\ref{EU1}) and (\ref{EU2}) can not be correct at that point. To see what happens at that point, one really needs to go back to (\ref{Lct}) and then one finds that (\ref{EU1}) is the right one to use when $d=2$ and $k=1$. 

More generally we have pole cancelations for all odd $k = 1,3,...,d-1$ where we shall use the exact counterterm (\ref{Lct}).

\subsection{The conformal anomaly on $S^1\times H^1$ revisited}\label{solved}
We will now compute the conformal anomaly on $S^1\times H^1$ boundary of $H^3$ using truncated counterterm renormalization. For $H^1$ we have two boundaries, and we have the coordinate range $\eta \in [-\eta_0,\eta_0]$. By taking this into account, the bulk and surface gravity actions become
\bea
I_{bulk} &=& \frac{1}{2G} \(r^2 - 1\) \ln\frac{1}{\delta}\cr
I_{surf} &=& - \frac{1}{G} \(r^2 - \frac{1}{2}\) \ln\frac{1}{\delta}
\eea
where $\delta = e^{-\eta_0}$. The truncated counterterm for a $d=2$ dimensional boundary is truncated at the first term because, at least naively, the second counterterm has a pole $1/(d-2)$ at $d=2$. Hence the truncated counterterm is given by
\bea
I_{trunc,ct} &=& \frac{1}{2G} \sqrt{r^2-1} r \ln\frac{1}{\delta}\cr
&=& \frac{1}{2G} \(r^2 - \frac{1}{2}\) \ln\frac{1}{\delta}
\eea
that originates from the truncated counterterm Lagrangian $\L_{ct} = d-1 = 1$ for $d=2$. Then 
\bea
I_{surf} + I_{trunc,ct} &=& \frac{1}{2G} \(-r^2+\frac{1}{2}\) + \O(1/r)
\eea
and the renormalized action becomes  
\bea
I_{trunc,ren} = I_{bulk} + I_{surf} + I_{trunc,ct} = - \frac{1}{4G} \ln\frac{1}{\delta}
\eea
This is the wrong answer.

The correct counterterm is given by 
\bea
I_{ct} &=& \frac{1}{4G} \(r^2-1\) \ln\frac{1}{\delta}
\eea
which corresponds to the counterterm Lagrangian 
\bea
\L_{ct} = \sqrt{1-\frac{1}{r^2}}
\eea
as one may infer directly from (\ref{Lct}) by taking $d=2$ and $k=1$ in which case we can use the ordinary hypergeomeric function since $k$ is odd. This hypergeometric function for $d=2$ and $k=1$ is equal to one, so the counterterm Lagrangian is easily computed from (\ref{Lct}). Using the correct counterterm, the conformal anomaly now comes out right as
\bea
I_{ren} &=& - \frac{1}{2G} \ln \frac{1}{\delta}
\eea
We know that this is the correct result since it matches the conformal anomaly on $S^2$ boundary.

It is important to notice that divergent terms, i.e. those that diverge as $r\rightarrow \infty$, are identical in $I_{trunc,ct}$ and $I_{ct}$. But there is another parameter $\delta$ and another divergence  $\eta_0 = \ln (1/\delta)$ in the term that does not diverge in $r$ that we would have otherwise called as a finite term. Here we can not tolerate this finite term in $r$, which is a logarithmically divergent term in $\delta$, to be regularization scheme dependent as that would give us the wrong value of the conformal anomaly.

\section{The Kounterterm method}\label{Kounterterm}
As has become clear, it is well-motivated to study alternative renormalization methods. The counterterm expansion (\ref{ct}) is known only for the first few terms and as we have seen, for some applications we need the exact counterterm to all orders, which is not known. 

One alternative is the Kounterterm renormalization method. In \cite{Anastasiou:2019ldc} it was shown that the Kounterterm is compatible with Dirichlet boundary condition for the metric. This suggests the Kounterterm method could be equivalent with the usual counterterm method although so far there is no general proof. 

The Kounterterm for a one-dimensional boundary is quite easy to understand and motivate. This is because in that case the counterterm in $d=1$ is necessarily very simple since the curvature of a one-dimensional boundary is zero, and also the first term in the counterterm expansion is proportional to $d-1 = 0$. So one should expect that the Kounterterm is equal to the surface term when $d=1$. Indeed this expectation turns out to be correct. From the general expression of the Kounterterm (\ref{KT}) below, we get
\bea
I_{Kt} = - \frac{1}{8\pi G} \int dx \sqrt{h} K = I_{surf}
\eea
when we put $d=1$.

The renormalized gravity action when using the Kounterterm method is given by
\bea
I_{ren} &=& I_{bulk} + I_{Kt}
\eea
In other words, the surface term is absent, and the Kounterterm is all there is. For general odd dimension $d$, the Kounterterm action is given by \cite{Anastasiou:2019ldc}
\ben
I_{Kt} &=& \frac{1}{8\pi G} \int d^d x \sqrt{h} \L_{Kt}\cr
\L_{Kt} &=& (-1)^{\frac{d+1}{2}} d \int_0^1 ds \delta^{i i_1\cdots i_{d-1}}_{j j_1\cdots j_{d-1}} K_{i}^{j} \(\frac{1}{2} R_{i_1 i_2}^{j_1 j_2} - s^2 K_{i_2}^{j_2} K_{i_3}^{j_3}\) \cdots \(\frac{1}{2} R_{i_{d-2} i_{d-1}}^{j_{d-2} j_{d-1}} - s^2 K_{i_{d-2}}^{j_{d-2}} K_{i_{d-1}}^{j_{d-1}}\)\cr
\delta_{i i_1\cdots i_{d-1}}^{j j_1\cdots i_{d-1}} &:=& \frac{1}{d!} \delta_i^j \delta_{i_1}^{j_1} \cdots \delta_{i_{d-1}}^{j_{d-1}} \pm \text{permutations of $ii_1\cdots i_d$}\label{KT}
\een
We consider $H^{d+1}$ foliated with $S^k \times H^{d-k}$ as 
\bea
ds^2 &=& d\rho^2 + \sinh^2\rho G_{ij} dx^i dx^j + \cosh^2\rho H_{ab} dx^a dx^b
\eea
where $G_{ij}$ and $H_{ab}$ denote the metrics of unit $S^k$ and unit $H^{d-k}$ respectively. For the boundary at $\rho = \rho_0$, we have 
\bea
R_{ij}^{kl} = \frac{2\delta_{ij}^{kl}}{\sinh^2\rho_0}, \qquad R_{ab}^{cd} = -\frac{2\delta_{ab}^{cd}}{\cosh^2\rho_0}
\eea
and 
\bea
K_i^j = \frac{\cosh\rho_0}{\sinh\rho_0} \delta_i^j, \qquad K_a^b = \frac{\sinh\rho_0}{\cosh\rho_0} \delta_a^b
\eea
We define $\L_{bulk}$ in the funny way as in (\ref{funny}) whose divergent part we have found to be given by
\bea
\L_{bulk,div,k} &=& \frac{r^k}{\(r^2-1\)^{k/2}} {}_2 F_1 \(-\frac{d}{2},\frac{1-k}{2},1-\frac{d}{2},\frac{1}{r^2}\)\cr
&=& 1 + \frac{d-2k}{2(d-2)} \frac{1}{r^2} + \frac{3d^2+8k^2-12dk+16k-6d}{8(d-2)(d-4)} \frac{1}{r^4} + \O\(\frac{1}{r^6}\)
\eea
where $r = \cosh\rho_0$. We define $\L_{Kt}$ by the same type of relation as
\bea
I_{Kt} &=& \frac{1}{8\pi G} \int d^d x \sqrt{h} \L_{Kt}
\eea
We want to cancel the divergences, so we want 
\bea
\L_{Kt} &=& - \L_{bulk,div}
\eea
and this is what we would now like to check. 

The Kounterterm simplifies when $d$ is odd and $k=0$ or $k=d$. For these cases we can explicitly carry out all the index contractions using 
\bea
1 &=& \delta^{i_1\cdots i_d}_{j_1\cdots i_d} \delta_{i_1}^{j_1} \delta_{i_2 i_3}^{j_2 j_3} \cdots \delta_{i_{d-1} i_d}^{j_{d-1} j_d}
\eea
and then we descend to the following two functions for the Kounterterm,
\ben
\L_{Kt,k=0}(d) &=& (-1)^{d} d \frac{\sinh\rho_0}{\cosh^d\rho_0} \int_0^1 ds \(1+s^2\sinh^2\rho_0\)^{\frac{d-1}{2}}\cr
\L_{Kt,k=d}(d) &=& (-1)^{\frac{d+1}{2}} d \frac{\cos\rho_0}{\sinh^d\rho_0} \int_0^1 ds \(1-s^2 \cosh^2\rho_0\)^{\frac{d-1}{2}}\label{Ktfn}
\een
These integrals can be evaluated for any complex-valued $d$ with the results
\ben
\L_{Kt,k=0} &=& (-1)^d d \frac{\sinh\rho_0}{\cosh^d\rho_0} {}_2 F_1 \(\frac{1}{2},\frac{1-d}{2},\frac{3}{2},-\sinh^2\rho_0\)\cr
\L_{Kt,k=d} &=& (-1)^{\frac{d+1}{2}} d \frac{\cos\rho_0}{\sinh^d\rho_0} {}_2 F_1 \(\frac{1}{2},\frac{1-d}{2},\frac{3}{2},\cosh^2\rho_0\)\label{Ktfn1}
\een
It can then be checked that 
\bea
\L_{bulk,div,0} + \L_{Kt,k=0} &=& 0\cr
\L_{bulk,div,d} + \L_{Kt,k=d} &=& 0
\eea
for any odd $d$. An important observation is that the Kounterterm does not destroy the univeral finite term, it only removes the divergences and leaves the finite term untouched. This also provides an independent confirmation that our guess for the infinite series expansion of the counterterm, for the case that the boundary is $S^d$, was correct. 

When $d$ is even, the Kounterterm is given by \cite{Anastasiou:2019ldc}
\bea
\L_{Kt} &=& \frac{(-1)^{d/2} d!}{2^{d-2} \[\(\frac{d}{2}-1\)!\]^2} \int_0^1 ds \int_0^s dt \delta^{i i_1\cdots i_{d-1}}_{j j_1\cdots i_{d-1}} \cr
&& K_{i}^{j} \(\frac{1}{2} R_{i_1 i_2}^{j_1 j_2} - s^2 K_{i_1}^{j_1} K_{i_2}^{j_2} + t^2 \delta_{i_1 i_2}^{j_1 j_2}\) \cdots \(\frac{1}{2} R_{i_{d-2} i_{d-1}}^{j_{d-2} j_{d-1}} - s^2 K_{i_{d-2}}^{j_{d-2}} K_{i_{d-1}}^{j_{d-1}} + t^2 \delta_{i_{d-2} i_{d-1}}^{j_{d-2} j_{d-1}}\)
\eea
For our boundaries we find the results
\bea
\L_{Kt,k=0} &=& \frac{(-1)^{d/2} d!}{2^{d-2} \[\(\frac{d}{2}-1\)!\]^2} \frac{\sinh\rho_0}{\cosh^{d-1}\rho_0}  \int_0^1 ds \int_0^s dt\(-1-s^2\sinh^2\rho_0 + t^2 \cosh^2\rho_0\)^{\frac{d}{2}-1}\cr
\L_{Kt,k=d} &=& \frac{(-1)^{d/2} d!}{2^{d-2} \[\(\frac{d}{2}-1\)!\]^2} \frac{\cosh\rho_0}{\sinh^{d-1}\rho_0}  \int_0^1 ds \int_0^s dt \(1-s^2 \cosh^2\rho_0 + t^2\sinh^2\rho_0\)^{\frac{d}{2}-1}
\eea
Evaluating the integral for $k=d=2,4,6$ gives the results 
\bea
\L_{Kt,k=d=2} &=& -1 + \O\(\frac{1}{r^2}\)\cr
\L_{Kt,k=d=4} &=& -1 + \frac{1}{r^2} + \O\(\frac{1}{r^4}\)\cr
\L_{Kt,k=d=6} &=& -1 + \frac{3}{4r^2} - \frac{3}{8r^4} + \O\(\frac{1}{r^6}\)
\eea 
The bulk action for $k=d=2,4,6$ gives us the following divergent terms
\bea
\L_{bulk,div,2} &=& 1\cr
\L_{bulk,div,4} &=& 1 - \frac{1}{r^2}\cr
\L_{bulk,div,6} &=& 1 - \frac{3}{4r^2} + \frac{3}{8r^4} 
\eea
Namely all these terms give rise to divergent terms in $I_{bulk,div,d}\sim r^d \L_{bulk,div,d}$. We see that the Kounterterm cancels all these divergences in the bulk action, which leaves us with a finite term plus a log divergent term. The finite term has no significance here as it can be absorbed into the log divergence. 

One may wonder why the formulas for the Kounterterm look so different when $d$ is even and odd. Of course these particular vector index contractions show that it does not work otherwise. However, we performed some studies using Mathematica by applying (\ref{Ktfn}) to the case when $d$ is even in the spirit of analytic continuation in a complex parameter $d$. This way we got a result for the Kounterterm that gave all the powerlaw divergences correctly, but we also got log divergences. We found the log divergence cancels the log divergence in the bulk action for $d=4$. But it seems to us the log divergences add up for $d=2$ and $d=6$ (we checked only up to $d=6$). One possibility could be that these signs could become wrong if Mathematica takes the wrong sign of some square roots.

\subsection{The finite term on $H^3$ revisited}\label{Fsolve}
We will now apply the Kounterterm method to compute the finite term for the $H^3$ foliation of $H^4$. This is one of those examples where we failed to compute the finite term by using the counterterm method, because for that we would need the infinite series expansion of the counterterm that is inaccessible to us. 

We have the bulk metric 
\bea
ds^2 &=& d\rho^2 + \cosh^2 \rho d\Xi_3^2
\eea
where $d\Xi_3^2 = d\eta^2 + \sinh^2\eta d\Omega_2^2$ is the metric of the boundary $H^3$. We get
\bea
\L_{bulk} &=& \frac{\sinh\rho_0}{\cosh^3\rho_0} \(\cosh^2\rho_0 + 2\)\cr
\L^{\rho=\rho_0}_{Kt} &=& - \frac{\sinh\rho_0}{\cosh^3\rho_0} \(\sinh^2\rho + 3\)
\eea
where we define $I_{bulk} := \frac{1}{8\pi G} \int_{H^3} d^3 x \sqrt{h} \L_{bulk}$ and the superscript $\rho=\rho_0$ means we compute the Kounterterm on the boundary surface $\rho=\rho_0$ which is $H^3$. Hence we get
\bea
\L_{bulk} + \L_{Kt}^{\rho=\rho_0} &=& 0
\eea
Let us now turn to the corner at $(\rho_0,\eta_0)$, which is where the two boundary components $\eta=\eta_0$ and $\rho=\rho_0$ meet. Let us expand the metric around the corner by defining local coordinates $x$ and $y$ as
\bea
\rho &=& \rho_0 + x\cr
\eta &=& \eta_0 + y \cosh\rho_0
\eea
Then we expand the bulk metric in the vicinity of the corner\footnote{There are actually two corners, located at $(\pm \rho_0,\eta_0)$.} as
\bea
ds^2 &=& dx^2 + dy^2 + \cosh^2 \rho_0 \sinh^2\eta_0 d\Omega_2^2 + ...
\eea
where $+...$ are higher order terms in a Taylor series expansion in $(x,y)$ where the corner is at $(x,y) = (0,0)$. Let us introduce polar coordinates $x + i y = r e^{i\phi}$. Then we regularize the corner which is most easily done by taking $r = \eps$ to be a constant.\footnote{We may also need to translate the origin by a vector $\eps e^{i\pi/4}$ but such a shift will not change the extrinsic curvature so we may be ignorant about where exactly the origin shall be located for the regularized corner surface.} Then we get the extrinsic curvature
\bea
K_{\phi}^{\phi} &=& \frac{1}{\eps}
\eea
We have the intrinsic curvature
\bea
R_{ij}{}^{kl} &=& \frac{2\delta_{ij}^{kl}}{\cosh^2\rho_0 \sinh^2\eta_0}
\eea
coming from the $S^2$ of radius $\cosh\rho_0\sinh\eta_0$, which is the corner submanifold. The measure factor (that is, the square root of the determinant of metric) of the corner is $\sqrt{h} = \eps \cosh^2\rho_0 \sinh^2\eta_0$. The Kounterterm Lagrangian from the corner is therefore
\bea
\L_{Kt}^{corn} &=& 3 \int_0^1 ds \delta_{\phi ij}^{\phi kl} K_{\phi}^{\phi} \frac{1}{2} R_{ij}{}^{kl}\cr
&=& \frac{1}{\eps \cosh^2\rho_0 \sinh^2\eta_0}
\eea
and the corresponding Kounterterm action is
\bea
I_{Kt}^{corn} &=& \frac{\Vol(S^2)}{8\pi G} \int_0^{\pi/2} d\phi \sqrt{h} \L_{Kt}^{corn}
\eea
Taking into account the fact that there are two corners at $(\pm\rho_0,\eta_0)$ and that we computed just the contribution form one of them above, we finally end up with the total corner contribution being 
\bea
I_{Kt}^{corn} &=& \frac{\pi}{2G}
\eea
which precisely agrees with (\ref{pufu}) that is computed by using $S^3$ foliation of $H^4$. It remains to understand what happens to the other boundary component and whether $I_{Kt}^{\eta=\eta_0} = 0$. We will not study this question here as the metric on that boundary component is rather complicated (see Appendix \ref{other}). It is rather clear that we already found the term that is responsible for the finite term.

To make this even more convincing, we may instead compute this corner term for generic odd $d$ and for simplicity, let us pick $k = d-1$. For $d=3$ this corresponds to a foliation with $S^2\times H^1$. The bulk metric is 
\bea
ds^2 &=& d\rho^2 + \sinh^2 d\Omega_{d-1}^2 + \cosh^2\rho d\eta^2
\eea
There are two corners at $(\rho_0,\pm \eta_0)$. We expand the metric around the corner as before and get
\bea
ds^2 &=& dx^2 + dy^2 + \sinh^2 \rho_0 d\Omega_{d-1}^2 + ...
\eea
and we regularize the corner as before by introducing polar coordinates as $r=\eps$ and $\phi \in [0,\pi/2]$ (and same for the other corner). The intrinsic curvature of the corner manifold is 
\bea
R_{ij}{}_{kl} &=& \frac{2\delta_{ij}^{kl}}{\sinh^2\rho_0}
\eea
and the extrinsic curvature of the corner has the only nonvanishing component 
\bea
K_{\phi}^{\phi} &=& \frac{1}{\eps}
\eea
We then get
\bea
\L_{Kt}^{corn} &=& (-1)^{\frac{d+1}{2}} d \int_0^1 ds \frac{1}{r} \frac{1}{d} \(\frac{1}{\sinh^2\rho_0}\)^{\frac{d-1}{2}}
\eea
where 
\bea
\frac{1}{d} &=& \delta_{\phi i_1\cdots i_{d-1}}^{\phi i_1\cdots i_{d-1}}
\eea
The corresponding action is 
\bea
I_{Kt}^{corn} &=& 2\times \frac{1}{8\pi G} \int_0^{\pi/2} d\phi \int d^d x \sqrt{h} \L_{Kt}^{corn} 
\eea
where the factor $2$ has been inserted taking into account the fact that there are two corners.  The measure factor is $\int d^d x \sqrt{h} = \Vol(S^{d-1}) \eps \sinh^{d-1}\rho_0$ canceling those factors from $\L_{Kt}$ leaving us with  
\bea
I_{Kt}^{corn} &=& (-1)^{\frac{d+1}{2}} \frac{\Vol(S^{d-1})}{8 G}
\eea
which, as one can easily see\footnote{To see this, we use the identity $\Gamma(-d/2)\Gamma(d/2+1) = - \pi/\sin(\pi d/2)$.}, is in exact agreement with the finite term (\ref{F}).

\section{The mass of a bulk geometry}\label{mass}
The on-shell gravity action is a function of the boundary metric $h_{\mu\nu}$ that we will  assume is timelike. One can use the on-shell action to define the quasilocal stress tensor on the boundary as
\ben
T^{\mu\nu}(x) &=& \frac{2}{\sqrt{-h}} \frac{\delta I(h)}{\delta h_{\mu\nu}(x)}\label{HJ}
\een
following \cite{Brown:1992br}. The original motivation for (\ref{HJ}) in \cite{Brown:1992br} came from an analogy with the Hamilton-Jacobi equation of a point particle
\bea
E &=& - \frac{\partial S}{\partial t}
\eea
where $E$ is the energy of the point particle, $S$ is the on-shell action as a function of boundary data (the position of the particle at some initial and final time), and $t$ is the final time. One may also notice the resemblance between the definition (\ref{HJ}) and the usual definition of the matter stress tensor in a gravity background. However, one should notice that the sign in the definition (\ref{HJ}). In a canonical stress tensor of some field $\phi$ that sign corresponds to having the canonical stress tensor
\bea
T^{\mu\nu} &=& - \(\frac{\partial \L}{\partial \partial_{\mu} \phi} \nabla^{\nu} \phi - g^{\mu\nu} \L\)
\eea
with an extra minus sign compared to what one usually has in the definition of the canonical stress tensor. 

Following \cite{Brown:1992br}, we assume that the timelike boundary metric can be put in the ADM form \cite{Arnowitt:1962hi}
\bea
ds^2 &=& - N^2 dt^2 + \sigma_{ab} \(dx^a + V^a dt\) \(dx^b + V^b dt\)
\eea
The timelike future pointing unit normal vector to constant time $t$ hypersurfaces is
\bea
u_t = -N,&&\qquad u_a = 0\cr
u^t=\frac{1}{N},&&\qquad u^a = \frac{V^a}{N}
\eea
Indeed tangent vectors on a constant $t$ hypersurfaces are $\partial_a$ corresponding to vectors $v^{\mu}$ with $v^t = 0$. Hence $v^{\mu} u_{\mu} = 0$ implies $u_a = 0$. It is future pointing because $u^t>0$. It is unit normalized, $u^{\mu} u_{\mu} = -1$. Let us assume that 
\ben
\nabla_{\mu} T^{\mu\nu} = 0\label{cons}
\een
on the boundary. This is not true in general for the quasilocal stress tensor if there is matter on the boundary. But we may assume the boundary is so far out at infinity that there is no matter there, and then we do have (\ref{cons}). Let us assume there is a Killing vector $\xi_{\mu}$ in the boundary surface. Then we can construct a conserved charge by considering the following integral over the boundary surface $B_d$ that we will assume extends from an initial constant time $t'$ hypersurface $C_{d-1}(t')$ to a final time $t''$ hypersurface $C_{d-1}(t'')$, 
\bea
0 &=& \int_{B_d} d^d x \sqrt{-h} \xi_{\mu} \nabla_{\nu} T^{\mu\nu} \cr
&=& \int_{B_d} d^d x\sqrt{-h} \nabla_{\nu}\(\xi_{\mu} T^{\mu\nu}\)\cr
&=& \int_{B_d} d^d x\partial_{\nu}\(\sqrt{-h}\xi_{\mu} T^{\mu\nu}\)\cr
&=& \[\int_{C_{d-1}} d^{d-1} x \sqrt{-h} \xi_{\mu} T^{\mu t}\]^{t''}_{t'}\cr
&=& -\[\int_{C_{d-1}} d^{d-1} x \sqrt{\sigma} \xi_{\mu} T^{\mu\nu} u_{\nu}\]^{t''}_{t'}
\eea
In the first line we used (\ref{cons}), in the second line we used $\nabla_{\mu} \xi + \nabla_{\nu} \xi_{\mu} = 0$ and in the last line we used $\sqrt{-h} = N \sqrt{\sigma}$ and $u_t = -N$. From this we conclude that there is a conserved charge \cite{Brown:1992br}
\bea
Q_{\xi} &=& -\int_{C_{d-1}} d^{d-1} x \sqrt{\sigma} \xi_{\mu} T^{\mu\nu} u_{\nu}
\eea
that does not change from initial to final time slices, but as those are arbitrary we conclude that $Q_{\xi}$ does not change over time. If $\lambda u_{\mu}$ is a Killing vector we have a candidate for a conserved mass,
\bea
M_{\lambda} &=& -\int_{C_{d-1}} d^{d-1} x \sqrt{\sigma} \lambda u_{\mu} T^{\mu\nu} u_{\nu}
\eea
We would now like to determine the coefficient function $\lambda$. In \cite{Brown:1992br} it is assumed that $u_{\mu}$ is a Killing vector\footnote{This condition is very restrictive. The timelike Killing vector is $C \partial_t$ with constant $C$. It corresponds to $u^t = C$, $u^a = 0$ and thus we must put $V^a = 0$.} and then we have a conserved quantity associated with $\lambda = 1$. However, this will not correspond to the canonical mass. Let us now determine $\lambda$ such that $M_{\lambda} = M$ is the canonical mass that corresponds to the Hamiltonian that generates time translations. We can fix the normalization by assuming a flat boundary metric 
\bea
ds^2 &=& - d t_c^2 + dx^a dx^a
\eea
We then define the canonical mass as
\bea
M_{t_c} &=& \int d^{d-1} x T^{t_c t_c} u_{t_c} u_{t_c}
\eea
because this exactly corresponds to the Hamiltonian that generates time translations. Here $u_{t_c} = -1$. Now the mass $M_{t_c}$ is defined with respect to a particular canonical time coordinate $t_c$, and the invariant quantity is $(t_c''-t_c') M_{t_c}$. If we change the metric to 
\bea
ds^2 &=& - N^2 dt^2 + dx^a dx^a
\eea
and define a new time coordinate $t=t_c/N$, we shall change the mass to 
\bea
M_t &=& N \int d^{d-1} x T^{tt} u_t u_t
\eea
where $u_t = - N$. This generalizes to the following definition of the mass 
\bea
M &=& \int_{C_{d-1}} d^{d-1} x \sqrt{\sigma} N u_{\mu} T^{\mu\nu} u_{\nu}
\eea
when $\xi_{\mu} = N u_{\mu}$ is a Killing vector. This agrees with the definition of mass in \cite{Balasubramanian:1999re}. We note that in the ADM coordinates $\xi^{t} = 1$ and $\xi^a = V^a$, and the condition that $\xi^{\mu}$ is a Killing vector on the boundary $B_d$ reduces to the condition that $V^a$ is a Killing vector on each constant time slice $C_{d-1}$ of the boundary.\footnote{This can be seen by expanding $\xi^{\lambda} \partial_{\lambda} h_{\mu\nu} + g_{\mu\lambda} \partial_{\beta} \xi^{\lambda} + g_{\nu\lambda} \partial_{\mu}\xi^{\lambda}$ using $\partial_t g_{\mu\nu} = 0$ and $\partial_{\mu} V^t = 0$.} This is a much less restrictive condition than what we get if we require $u^{\mu}$ is a Killing vector, which puts $V^a = 0$.

For a Lorentzian bulk and boundary (we will not discuss spacelike initial and final boundaries) the gravity action is given by $I = I_{bulk} + I_{surf} + I_{ct}$ where
\bea
I_{bulk} &=& \frac{1}{16\pi G} \int d^{d+1} x \sqrt{-g} \(R - 2\Lambda\)\cr
I_{surf} &=& \frac{1}{8\pi G} \int d^d x \sqrt{-h} K\cr
I_{ct} &=& \frac{1}{8\pi G} \int d^d x \sqrt{-h} \L_{ct}
\eea
Here $\L_{ct}$ is an expression contructed out of the boundary metric and its derivatives. If we let $n_{\mu}$ denote the unit normalized outward pointing normal vector to the timelike boundary,
\ben
g^{\mu\nu} n_{\mu} n_{\nu} &=& 1\label{unit}
\een
then the boundary metric can be expressed as
\bea
h_{\mu\nu} &=& g_{\mu\nu} - n_{\mu} n_{\nu}
\eea
where $g_{\mu\nu}$ is the bulk metric. We define the projector onto the boundary as
\bea
h_{\mu}^{\nu} &=& \delta_{\mu}^{\nu} - n_{\mu} n^{\nu}
\eea
where all indices are rised by $g^{\mu\nu}$. In fact $h_{\mu\nu}$ has no inverse. But we  may define $h^{\mu\nu} := g^{\mu\kappa} g^{\nu\tau} h_{\kappa\tau}$. The exterior curvature of the boundary is defined as
\bea
K_{\mu\nu} &=& h_{\mu}^{\kappa} h_{\nu}^{\tau} \nabla_{\kappa} n_{\tau}
\eea
By using (\ref{unit}) we then get
\bea
K_{\mu\nu} &=& \nabla_{\mu} n_{\nu} - n_{\mu} n^{\tau} \nabla_{\tau} n_{\nu}
\eea
and its trace
\bea
K = g^{\mu\nu} K_{\mu\nu} = \nabla^\mu n_{\mu}
\eea
The variation of the on-shell action is 
\bea
\delta I_{cl} &=& \int d^d x \pi^{\mu\nu} \delta h_{\mu\nu} 
\eea
where
\bea
\pi^{\mu\nu} &=& - \frac{1}{16\pi G} \sqrt{-h} \(K^{\mu\nu} - h^{\mu\nu} K\)
\eea
Comparing with (\ref{HJ}) we see that 
\bea
T^{\mu\nu}_{bulk} = \frac{2}{\sqrt{-h}} \pi^{\mu\nu} = - \frac{1}{8\pi G} \(K^{\mu\nu} - h^{\mu\nu} K\)
\eea
This contribution to the stress tensor comes entirely from the bulk action. However, this bulk stress tensor is divergent and needs to be renormalized. We do that by adding the contribution coming from the counterterm, and define the renormalized stress tensor \cite{Balasubramanian:1999re} as
\bea
T^{\mu\nu}_{ren} &=& T^{\mu\nu}_{bulk} + T^{\mu\nu}_{ct}
\eea
where
\bea
T^{\mu\nu}_{ct} &=& \frac{2}{\sqrt{-h}} \frac{\delta I_{ct}}{\delta h_{\mu\nu}}
\eea

The boundary metric can be expressed as
\bea
h_{\mu\nu} &=& \sigma_{\mu\nu} - u_{\mu} u_{\nu}
\eea
Using this, one finds that the unrenormalized bulk mass becomes
\bea
M_{bulk} &=& - \frac{1}{8\pi G} \int d^{d-1} x \sqrt{\sigma} N K_{C}
\eea
where
\bea
K_C &=& \sigma^{ab} K_{ab} 
\eea
Here we define $K_{ab}$ as the pullback to a constant time slice of the extrinsic curvature $K_{\mu\nu}$ of the boundary as an embedded surface in the bulk. The counterterm Lagrangian gives an additional contribution to the mass
\bea
M_{ct} &=& \int_{C_{d-1}} d^{d-1} x \sqrt{\sigma} N u_{\mu} T^{\mu\nu}_{ct} u_{\nu}
\eea
The counterterm stress tensor can of course be computed by a metric variation of the counterterm action using  (\ref{HJ}), but it can also be computed using the canonical formalism as
\bea
T^{\mu\nu}_{ct} &=& - \(\frac{\partial \L_{ct}}{\partial \partial_{\mu} \phi} \nabla^{\nu} \phi - h^{\mu\nu} \L_{ct}\)
\eea
where $\phi$ runs over all the fields, which are comprised of the boundary metric and its derivatives (in the absence of matter), in the counterterm Lagrangian. Now if $\xi_{\mu} = N u_{\mu}$ is a timelike Killing vector, then it must be proportional to $\partial_t$ and then we have $\xi^{\nu} \nabla_{\nu} \phi \sim \partial_t \phi = \L_\xi \phi = 0$, which is just saying that time translation is an isometry. That means that we have
\bea
T^{\mu\nu} \xi_{\nu} &=& \xi^{\mu} \L_{ct}
\eea
and by using this we get
\ben
M_{ct} &=& - \int_{C_{d-1}} d^{d-1} x \sqrt{\sigma} N \L_{ct}\label{Lag}
\een
The minus sign comes out right by using $u^{\mu} u_{\mu} = -1$. In the end the unusual sign convention of the Brown-York quasilocal stress tensor is just a convention. No physical quantity depends on this convention. We thus conclude is that the mass contribution that comes from the counterterm Lagrangian is simply equal to minus that counterterm Lagrangian,
\bea
M_{ct} &=& - L_{ct}
\eea
when the metric has a timelike Killing vector.

\subsection{The mass of AdS}
Let us illustrate this by computing the mass of $AdS_{d+1}$ with the timelike boundary $B_d = \mb{R} \times H^{d-1}$. The bulk metric is 
\bea
ds^2 &=& - \sinh^2 \rho dt^2 + d\rho^2 + \cosh^2 \rho H_{ab} dx^a dx^b
\eea
The boundary $B_d$ is at some constant cutoff value $\rho = \rho_0$ with the outward pointing unit normal $n^{\rho}=1$. The metric on a constant time slice $C_{d-1}$ on the boundary is $h_{ab} = \cosh^2 \rho_0 H_{ab}$. We then get
\bea
K_{ab} = \frac{1}{2} \(\nabla_a n_b + \nabla_b n_a\) = \sinh\rho_0 \cosh\rho_0 H_{ab}
\eea
and 
\bea
K_C = \sigma^{ab} K_{ab} = (d-1) \frac{\sinh\rho_0}{\cosh\rho_0}
\eea
Here the lapse function is $N=\sinh\rho_0$ so the bulk mass becomes 
\bea
M_{bulk} = - \frac{\Vol(H^{d-1})}{8\pi G} (d-1) \cosh^{d-2}\rho_0 \sinh^2\rho_0
\eea
To this we add the contribution from the counterterm, which is 
\bea
M_{ct} &=& - L_{ct}
\eea
The counterterm Lagrangian was obtained in (\ref{EU1}) Euclidean signature for foliation of $H^{d+1}$ by $S^1\times H^{d-1}$. This result carries over to Lorentzian signature with AdS$_{d+1}$ foliated by $\mb{R} \times H^{d-1}$ with an additional minus sign,
\bea
L_{ct} &=& \frac{1}{8\pi G} \int d^{d-1} x \sqrt{\sigma} N \L_{ct}\cr
\L_{ct} &=& - (d-1) \sqrt{1-\frac{1}{r^2}}
\eea
Here $r = \cosh\rho_0$. It is now easy to see that the counterterm mass cancels the bulk mass. The renormalized mass of AdS$_{d+1}$, at least for this choice of boundary, is zero,
\bea
M_{ren} = M_{bulk} + M_{ct} = 0
\eea
But we may also compute the mass of AdS$_{d+1}$ using the foliation with boundary $B_d = \mb{R}\times S^{d-1}$. The metric is
\ben
ds^2 &=& - \cosh^2 \rho + d\rho^2 + \sinh^2\rho d\Omega_{d-1}^2\label{spherefol}
\een
We get
\bea
M_{bulk} &=&  - \frac{\Vol(S^{d-1})}{8\pi G} (d-1) \sinh^{d-2} \rho_0 \cosh^2 \rho_0
\eea
and from (\ref{EU2}) we deduce that 
\bea
\L_{ct} &=& - (d-1) \sqrt{1+\frac{1}{s^2}}
\eea
where $s= \sinh\rho_0$. Again we find that 
\bea
M_{ren} = M_{bulk} + M_{ct} = 0
\eea
If we were to compute the mass of $AdS_{d+1}$ using background subtraction, we would proceed by choosing the background as a space determined by the asymptotic geometry, which in this case is $AdS_{d+1}$. We would next pick a boundary in this background that has as its induced metric the same metric as the boundary metric of the original space, but since the original space is again $AdS_{d+1}$, the boundary of the reference space and the original space will be the same. Then we compute $T^{\mu\nu}_{ref}$ of the reference space and find that this is identical with $T^{\mu\nu}_{bulk}$ of the original space and finally we subtract (background subtraction) to get the result
\bea
M_{ren} = M_{bulk} - M_{ref} = 0
\eea
Hence background subtraction gives the same result as the counterterm method.

We think it is reassuring that we get the same renormalized mass for either choice of boundary. This is in accordance with our general philosophy that the boundary is just a regulator surface and the renormalized bulk mass should not depend on the choice of this boundary, as long as it is taken towards infinity, or the regulator cutoff $\eps$ is taken to zero.

However, we also notice that in the literature a different renormalized mass of AdS is presented. One particularly interesting result is the mass of AdS$_5$ that was shown to match precisely with the Casimir energy of the dual 4d $N=4$ SYM. All these results are obtained using a truncated counterterm series expansion. We summarize these results in appendix \ref{mass}. 

We notice that the mass of AdS may be computed in a different way as the on-shell action of thermal AdS by taking the zero temperature limit or $\beta$ to infinity. In that limit the partition function of the dual CFT is dominated by the Casimir energy $E$ as $Z \sim e^{-\beta E}$ and by AdS/CFT this is idenfied with the exponent of minus the on-shell gravity action $e^{-I}$. This might help us to understand what goes wrong when one tries to identify the Casimir energy of the dual CFT with the mass of AdS. Namely by this chain of reasoning we might need to study thermal AdS at some intermediate step, even if we take the zero temperature limit in the end. One may view thermal AdS as a regulator. We would now like to argue that this might be a bad regulator and that this could be the reason we can not use it, despite we take the zero temperature limit in the end. If the regulator is bad, then no matter we take the limit in the end or not, we will get the wrong result. Thermal AdS is obtained by imposing a periodic identification on time $t$ so that the boundary changes from $H^1 \times S^{d-1}$ into $S^1_{\beta} \times S^{d-1}$. For the relevant AdS/CFT applications that we have in mind, we also need to be concerned with supersymmetry. To preserve some supersymmetry we need to turn on the time component of a background gauge field $A_t$ in AdS. Such a gauge field originates from a graviphoton field in the metric on AdS times a sphere, where it appears in the metric on the form $d\mu^2 + \mu^2 \(d\phi + A_t dt\)^2$ for some radial coordinate $\mu$ and angle coordinate $\phi$ on the sphere.\footnote{There is one $\mu$ and one $\phi$ for each Cartan of the isometry group of the sphere and we need to sum over them to get the full metric.} To preserve some supersymmetry, one may want to turn on a constant gauge field $A_t$ along the time direction. But that is incompatible with the periodic identification of time coordinate $t$. To see this, we define a new coordinate as $\phi' := \phi + A_t t$. If $t$ is periodic, this new coordinate $\phi'$ will in general not have the same periodicity as the original angle $\phi$. So turning on a constant $A_t$ will create a conical singularity in the bulk metric at $\mu = 0$ where the classical gravity description breaks down.

In \cite{Bak:2019lye} we presented an alternative way to compute the Casimir energy on $\mb{R}\times S^5$ by relating this to the conformal anomaly of a conical deformation of $S^d$ that in turn is conformally related to $S^1 \times H^{d-1}$ which is the boundary of AdS as opposed to thermal AdS. Here things are under much better control. There is a black hole solution that is a deformation away from AdS in which one can turn on background gauge fields. The resulting Casimir energy and the corresponding match with the dual CFT in \cite{Bak:2019lye} relied on assuming that the mass of the black hole in AdS$_7$ has a certain zero point value. Interestingly, that zero point value of the mass is what we will obtain below precisely when we compute the mass using the untruncated counterterm. We think this provides strong evidence that the untrunctated counterterm is really the correct counterterm to use.

\subsection{The mass of a black hole in AdS$_7$}
To get a nonzero renormalized mass we may put a black hole into the AdS bulk that deforms the interior geometry while keeping the asymptotic AdS geometry unchanged. Let us consider a two-charged black hole in AdS$_7$ \cite{Cvetic:1999xp}. Its mass was obtained in \cite{Bak:2019lye} by partly using holographic renormalization and partly using background subtraction to fix the zero point of the energy. It was observed in \cite{Bak:2019lye} that holographic renormalization did not give the correct zero point of the energy and for that purpose a constant shift of the energy was introduced by hand. Here we will see this problem can be avoided if we do not truncate the counterterm. Let us begin with the background subtraction method. Then to compute the black hole mass, we subtract the unrenormalized mass $M_{ref}$ of AdS$_7$. Using the same notations as in \cite{Bak:2019lye}, we define
\bea
f &=& r^2 H_1 H_2 - 1 - \frac{m}{r^4}\cr
H_1 &=& 1 + \frac{q_1}{r^4}\cr
H_2 &=& 1 + \frac{q_2}{r^4}
\eea
where $q_1,q_2$ are two charges of the black hole and $m$ is a mass parameter of the black hole. The metric of the black hole is given by
\bea
ds^2 &=& - A^2 dt^2 + \frac{dr^2}{C^2} + B^2 d\Xi_5^2 
\eea
where $d\Xi_5^2$ denotes the metric on unit $H^5$ and 
\bea
A &=& (H_1 H_2)^{-2/5} \sqrt{f}\cr
B &=& (H_1 H_2)^{1/10} r\cr
C &=& (H_1 H_2)^{-1/10} \sqrt{f}
\eea
Now there is a natural candidate for a reference background geometry, namely the geometry that we get by putting $q_1=q_2=m=0$. The metric then becomes the above with
\bea
A &=& \sqrt{r^2-1}\cr
B &=& r\cr
C &=& \sqrt{r^2-1}
\eea 
which gives the metric of AdS$_7$. Let us start by computing the mass using background subtraction. Background subtraction has the advantage that we do not need to compute counterterms. All we need to know is the formula for the bulk stress tensor,
\bea
T_{bulk,\mu\nu} &=& - \frac{1}{8\pi G} \(K_{\mu\nu} - h_{\mu\nu} K\)
\eea
For the black hole solution we get
\bea
T_{bulk,tt} &=& - \frac{1}{8\pi G} \(-C A A' - (- A^2) C \(\frac{A'}{A} + \frac{5 B'}{B}\)\)\cr
&=& - \frac{1}{8\pi G} \frac{5 A^2 C B'}{B}
\eea
The corresponding mass as measured by the canonical boundary time $t_c=A t$ is obtained by integrating over space and dividing by $A^2$,
\bea
M_{bulk} &=& \frac{1}{A^2} \int d^5 x \sqrt{h} T_{bulk,tt}
\eea
Here $\sqrt{h} = B^5$ so we get
\bea
M_{bulk} &=& \frac{\Vol(H^5)}{8\pi G} \M_{bulk}\cr
\M_{bulk} &=& - 5 C B^4 B'
\eea
We get
\bea
\M_{bulk} &=& - 5r^5 + \frac{5}{2}r^3 + \frac{5}{8}\(1-4(q_1+q_2)\)r + \(\frac{5}{16} + \frac{5}{2} m  - \frac{5}{4}\(q_1+q_2\) \) \frac{1}{r} + \O\(\frac{1}{r^2}\)
\eea
We now compute the corresponding quantity for the reference space AdS$_7$ where we keep the same boundary metric. Thus we take the bulk metric as
\bea
ds^2 &=& - A^2_b dt^2 + \frac{dr^2}{C_b^2} + B_b^2 d\Xi_5^2 
\eea
with
\bea
A_b &=& \sqrt{r^2-1}\cr
B_b &=& r\cr
C_b &=& \sqrt{r^2-1}
\eea 
and then we identify the boundary at some $r=r_b$ for which the boundary metric 
\bea
ds^2 &=& - \(r_b^2-1\) dt^2 + r_b^2 d\Xi_5^2
\eea
becomes the same as above. This way we conclude that $r_b=B(r)$ and then the mass with respect to the canonical boundary time $t_c = t \sqrt{r^2_b-1}$ becomes
\bea
M_{ref} &=& \frac{\Vol(H^5)}{8\pi G} \M_{ref}\cr
\M_{ref} &=& - r_b^5 \sqrt{1-\frac{1}{r_b^2}}
\eea
for which we have the $1/r$ expansion
\bea
\M_{ref} &=& - 5 r^2 + \frac{5}{2} r^3 + \frac{5}{8} \(1-4(q_1+q_2)\) r + \(\frac{5}{16} + \frac{3}{4} \(q_1+q_2\) \) \frac{1}{r} + \O\(\frac{1}{r}\)
\eea
The renormalized mass, which we identify as mass of the black hole, is given by the difference \cite{Cvetic:1999ne} 
\bea
M_{ren} = M_{bulk} - M_{ref} = \frac{\Vol(H^5)}{8\pi G} \(\frac{5m}{2} - 2\(q_1+q_2\)\)
\eea
Let us now compute this mass by using holographic renormalization. The advantage here is that we do not need to introduce a background geometry. Instead we shall compute a counterterm that is made up of curvature invariants of the boundary metric 
\bea
ds^2 &=& - A^2 dt^2 + B^2 d\Xi_5^2 
\eea
The counterterm does not feel the bulk geometry, not even the vicinity of the bulk geometry near the boundary. The counterterm is only a function of the boundary metric. Moreover, it is universal, the same for any bulk space that shares the same boundary. We have obtained the counterterm when we foliated $H^7$ with $S^1 \times H^5$ in (\ref{EU1}) and that result can be taken over here to give us
\bea
\L_{ct} &=& 5 \sqrt{1-\frac{1}{B^2}}
\eea
The Lagrangian is the space integral 
\bea
L_{ct} &=& \frac{\Vol(H^5)}{8\pi G} 5 B^5 \sqrt{1-\frac{1}{B^2}}
\eea
This contributes to the mass as computed with respect to the boundary time $t_c = A t$ that is simply $M_{ct} = - L_{ct}$. We now see that we recover the result of background subtraction. The counterterm mass is exactly equal to minus the mass of the reference background geometry,
\bea
M_{ct} &=& -M_{ref}
\eea

\section{Another formula for the counterterm}\label{formula}
By combining this result with our earlier result $L_{ct} = - M_{ct}$, we are led to conjecture the following quite general formula for the counterterm Lagrangian,
\ben
L_{ct} &=& - M_{ref}\label{MMA}
\een
where
\bea
M_{ref} &=& - \frac{1}{8\pi G} \int d^{d-1} x \sqrt{\sigma} N (K_b)_C
\eea
where $(K_b)_C := \sigma^{ab} (K_b)_{ab}$ denotes the extrinsic curvature computed when the boundary metric is kept fixed but the boundary surface is embedded in the reference AdS space. It is not always possible to find such a surface such that its metric as induced from the reference AdS space coincides with the original boundary metric. But for those cases we may change the definition of $(K_b)_{ab}$ slightly, following the same idea as Mann and Marolf \cite{Mann:2005yr} used in a flat background, and use the Gauss-Codazzi equation in an AdS background 
\ben
R_{\mu\nu\lambda\rho} &=& \t R_{\mu\nu\lambda\rho} + K_{\mu\lambda} K_{\nu\rho} - K_{\nu\lambda} K_{\mu\rho}\cr
\t R_{\mu\nu\lambda\rho} &=& - \(g_{\mu\lambda} g_{\nu\rho} - g_{\nu\lambda} g_{\mu\rho}\)\label{GC}
\een
as an implicit definition of $(K_b)_{\mu\nu}$ in terms of the intrinsic curvature $R_{\mu\nu\lambda\rho}$ on the boundary, from which we get $(K_b)_{ab}$ as its pullback to a constant time slice. Here $\t R_{\mu\nu\lambda\rho}$ is the curvature of AdS. With such an implicit definition we have now $(K_b)_{ab}$ as an (implictly defined) function of the intrinsic curvature of the boundary. This means that the formula (\ref{MMA}) now extends to cases when there is no embedding in AdS that gives the boundary metric, just using the implicit form of $(K_b)_{ab}$ all we need to compute (\ref{MMA}) is the boundary metric itself. With such an understanding, (\ref{MMA}) becomes quite general and applies to any bulk geometry that is asymptotically AdS where the boundary surface is such that it has a timelike Killing vector. 

It would be better if we could solve (\ref{GC}) and get the explicit expression for (\ref{MMA}). Gravity simplifies at large $d$ \cite{Emparan:2013moa}. For large dimension $d$ we can try to solve (\ref{GC}) by making an expansion in $1/d$. Contracting (\ref{GC}) with the inverse boundary metric $h^{\mu\nu}$, we get
\bea
R &=& - \frac{d(d-1)}{\l^2} + K^2 - K_{\mu\nu} K^{\mu\nu}
\eea
where we introduced the AdS radius $\l$ and where $K := h^{\mu\nu} K_{\mu\nu}$. To leading order in $1/d$ we can neglect the last term, and we can approximate $K$ by $K_C$. Then we can solve for $K$ as
\ben
K &=& \sqrt{R + \frac{d^2}{\l^2}} \(1+\O\(\frac{1}{d}\)\)\label{A1}
\een
This result is exact in $\l$ to leading order in $1/d$. As a consistency check, when $\l$ small, we have the small-$\l$ and large-$d$expansion 
\bea
K &=& \frac{d}{\l} + \frac{l}{2d} R + ...
\eea
which agrees with the counterterm Lagrangian as an $\l$ expansion to leading order in $1/d$. It is also easy to get an improvement of (\ref{A1}) by matching with its known $\l$ expansion that is exact in $d$
\ben
K &=& \frac{d-1}{\l} + \frac{\l}{2(d-1)} R + \O\(\l^3\)\label{small}
\een
This way we are led to the formula
\ben
K &=& \sqrt{\frac{d-1}{d-2} R + \frac{(d-1)^2}{\l^2}}\label{before}
\een
whose small $\l$ expansion agrees with (\ref{small}) up to order $\l$. The formula (\ref{before}) we have seen before, in (ref{here}). To get a formula that works for higher orders in $\l$ we need to find a closed formula for $K$ as a function of the full Riemann tensor of the boundary metric as well as of all its derivatives. This problem might be possible to study systematically as a $1/d$ expansion. 
 
The problem of inverting (\ref{GC}) simplifies in the flat space limit $\l\rightarrow \infty$ where the problem reduces to inverting the Gauss-Codazzi equations \cite{Mann:2005yr}
\ben
R_{\mu\nu} &=& K_{\mu\nu} K - K_{\mu}^{\lambda} K_{\lambda\nu}\label{GCFa}\\
R &=& K^2 - K_{\mu\nu} K^{\mu\nu}\label{GCFb}
\een
This problem was solved when $d=3$ \cite{Visser:2008gx}, but the method used there does not generalize to other dimensions. For other dimensions we can instead make a $1/d$ expansion to arbirary order. We have the following leading scaling behavior with $d$ for the various fields,
\bea
R_{\mu\nu\lambda\kappa} &\sim & 1\cr
R_{\mu\nu} &\sim & d\cr
R &\sim & d^2\cr
h_{\mu\nu} &\sim & 1
\eea
From (\ref{GCFa}) we deduce that
\bea
K_{\mu\nu} &\sim & 1\cr
K &\sim & d
\eea
This implies that in the large $d$ limit $K_{\mu\nu} \sim h_{\mu\nu}$ since that is the only way that we could get $K \sim h^{\mu}_{\mu} = d$. Then we also get $K_{\mu}^{\lambda} K_{\lambda\nu} \sim  1$ and $K_{\mu\nu} K^{\mu\nu} \sim d << K^2 \sim d^2$. With these preparations, we can solve the Gauss-Codazzi equations to leading order in $1/d$,
\ben
K &=& \sqrt{R}\cr
K_{\mu\nu} &=& \frac{R_{\mu\nu}}{\sqrt{R}}\label{initial}
\een
where we chose inward pointing normal vector to get positive signs, $K>0$, when $R>0$. We will not address the question what happens when $R<0$. We now get the $1/d$ expansion by iterating 
\bea
K_{\text{new},\mu\nu} &=& \frac{1}{K} \(R_{\mu\nu} + K_{\mu\lambda} K^\lambda{}_\nu\)\cr
K_{\text{new}} &=& \sqrt{R + K_{\mu\nu} K^{\mu\nu}}
\eea
with the initial conditions (\ref{initial}). We can see how each iteration corrects terms one by one in a power series expansion in $1/d$ by looking at an example where we know the exact result. Let us write the flat reference space bulk metric in the form 
\bea
ds^2 &=& -dt^2 + dr^2 + r^2 d\Omega_{d-1}^2
\eea
and let the boundary surface we the $\mb{R}\times S^{d-1}$ located at a constant large $r$. Then on a constant time slice $S^{d-1}$, we have
\ben
R_{ij} &=& \frac{h_{ij} (d-2)}{r^2}\cr
K_{ij} &=& r G_{ij}\cr
K &=& \frac{d-1}{r}\label{example1}
\een
for inward pointing unit normal. Here $d\Omega_{d-1}^2 = G_{ij}dx^i dx^j$ and $h_{ij} = r^2 G_{ij}$. One may easily check that these satisfy the Gauss-Codazzi equation $R_{ij} = K_{ij} K - h^{kl} K_{ik} K_{jl}$. Now let us solve this equation iteratively. Then we get 
\bea
K &=& \frac{d}{r} \(1 - \frac{3}{2d} - \frac{1}{8d^2} - \frac{3}{16d^3} + ... \)\cr
K &=& \frac{d}{r} \(1 - \frac{1}{d} - \frac{1}{2d^2} - \frac{1}{2d^3} + ...\)\cr
K &=& \frac{d}{r} \(1 - \frac{1}{d} + \frac{0}{d^2} - \frac{1}{2d^3} +...\)
\eea
and we can see that by each iteration we increase the number of correct coefficients by one order in the $1/d$ expansion such that we approach the exact result $K = \frac{d-1}{r} = \frac{d}{r} \(1 - \frac{1}{d}\)$.

We can alternatively make a general ansatz for the infinite $1/d$ expansion. It is now conventient to change our notation and let $R$ and $K$ represent the matrices $R_{\mu\nu}$ and $K_{\mu\nu}$. We write $\tr R$ and $\tr K$ for $R_{\mu}^{\mu}$ and $K_{\mu}^{\mu}$. We introduce coefficients of the $1/d$ expansion that we denote as $a^{(n)}$ where $n$ is associated with the order in $1/d$. These coefficients are matrices. We use the notation $\tr R^2 = R_{\mu\nu} R^{\mu\nu}$ and $\tr^2 R = \(R_{\mu}^{\mu}\)^2$. Our ansatz for the $1/d$ expansion is 
\bea
K &=& \frac{1}{\sqrt{\tr R}} \sum_{n=1}^{\infty} \frac{a^{(n)}}{\(\tr R\)^{n-1}}
\eea
Here the various quantities scale with the dimension $d$ as
\bea
R &\sim & d\cr
\tr R &\sim & d^2\cr
\tr R^n &\sim & d^{n+1}\cr
a^{(n)} &\sim & d^n\cr
\tr a^{(n)} &\sim & d^{n+1}
\eea
Then we solve the Gauss-Codazzi equation
\bea
K \tr K - K^2 &=& R
\eea
with respect to $K$ as given by the above ansatz. By organizing the expansions in powers of $1/d$, we end up with the following relations among the coefficient matrices,
\bea
a^{(q)} + \frac{a^{(1)} \tr a^{(q)}}{\tr R} + ... + \frac{a^{(q-1)} \tr a^{(1)}}{\tr R} - a^{(1)} a^{(q-1)} -... - a^{(q-1)} a^{(1)} &=& 0
\eea
for $q=2,3,4,...$ and we start the iteration by declaring the initial condition
\bea
a^{(1)} &=& R
\eea
We may solve for $a^{(q)}$ as
\bea
a^{(q)} &=& \sum_{n=1}^{q-1} \(a^{(n)} a^{(q-n)} - \frac{R}{2\tr R} \tr\( a^{(n)} a^{(q-n)}\)\)\cr
&& + \sum_{n=2}^{q-1} \(\frac{R}{2(\tr R)^2} \tr \(a^{(n)}\) \tr \(a^{(q-n+1)}\) -\frac{1}{\tr R} a^{(n)} \tr \(a^{(q-n+1)}\)\)
\eea
By the first iteration we get
\bea
a^{(2)} &=& R^2 - \frac{R\tr(R^2)}{2\tr R}
\eea
We would now like to obtain $\tr K$ that will be our counterterm Lagrangian. If we introduce a fictious parameter $\eps=1$ that keeps track of the order $n$ in the $1/d$ expansion, then we have
\bea
\tr K &=& \sqrt{\tr R} \sum_{n=1}^{\infty} c_n \eps^{n}
\eea
where the coefficients are
\bea
c_{n-1} &=&  \frac{\tr(a^{(n)})}{\tr^n R}
\eea
and can be computed using the above iterative formula with the results
\bea
c_0 &=& 1\cr
c_1 &=& \frac{\tr R^2}{2\tr^2 R}\cr
c_2 &=& \frac{\tr R^3}{\tr^3 R} - \frac{5 \tr^2 R^2}{8 \tr^4 R}\cr
c_3 &=& \frac{5\tr R^3}{2\tr^3 R} - \frac{7 \tr R^2 \tr R^3}{2\tr^5 R} + \frac{21 \tr^3 R^2}{16 \tr^6 R}\cr
c_4 &=& \frac{7\tr R^5}{\tr^5 R} - \frac{45 \tr R^2 \tr R^4}{4\tr^6 R} - \frac{9 \tr^2 R^3}{2\tr^6 R} + \frac{99 \tr R^3 \tr^2 R^2}{8 \tr^7 R} - \frac{429 \tr^4 R^2}{128 \tr^8 R}
\eea
We were hoping to be able to guess the exact formula from the $1/d$ expansion. This hope did not get realized. This expansion looks rather random. The coefficients may get somewhat nicer when we compute the square quantity
\ben
\(\tr K\)^2 &=& \tr R \sum_{n=1}^{\infty} C_n \eps^{n}\label{EX}
\een
Then we find the coefficients
\bea
C_0 &=& 1\cr
C_1 &=& \frac{\tr R^2}{\tr^2 R}\cr
C_2 &=& \frac{2 \tr R^3}{\tr^3 R} - \frac{\tr^2 R^2}{\tr^4}\cr
C_3 &=& \frac{5 \tr R^4}{\tr^4 R} + \frac{2 \tr^3 R^2}{\tr^6 R} - \frac{6 \tr R^2 \tr R^3}{\tr^5 R }\cr
C_4 &=& \frac{14 \tr R^5}{\tr^5 R} - \frac{20 \tr R^2 \tr R^4}{\tr R^6} - \frac{8 \tr^2 R^3}{\tr^6 R} + \frac{20 \tr^2 R^2 \tr R^3}{\tr^7 R} - \frac{5 \tr^4 R^2}{\tr^8 R}
\eea
but still it looks rather random. We can make progress with exact formulas for some special boundaries though. Let us begin by assuming the boundary is $\mb{R} \times S^{d-1}$. Then we have
\bea
\tr R^n &=& \frac{1}{r^{n+1}}(d-2)^n (d-1)
\eea
The above expansion (\ref{EX}) gives  
\bea
\(\tr K\)^2 &=& \tr R \(1 + \frac{\eps}{d-1} + \frac{\eps^2}{(d-1)^2} + ...\) \cr
&=& \frac{(d-1)(d-2)}{r^2} \frac{d-1}{d-2}\cr
&=& \(\frac{d-1}{r}\)^2
\eea
Taking the square root, we get (\ref{example1}). 

Let us next consider the generalization to the boundary $S^k\times \mb{R}^{d-k}$. Then we get
\bea
\tr R^n &=& \frac{1}{r^{n+1}}(k-1)^n k
\eea
The above expansion (\ref{EX}) now gives
\bea
(\tr K)^2 = (\tr R) \frac{k}{k-1} = \(\frac{k}{r}\)^2
\eea
Taking the square root, we get $\tr K = k/r$, which is the correct answer for the trace of the extrinsic curvature for this boundary \cite{Kraus:1999di}. 

We also made the following anecdotal observation. If we assume that
\ben
\tr R^n &=& \tr^n R\label{oned}
\een
then the series expansion becomes 
\bea
\(\tr K\)^2 &=& \tr R \(1 + \eps + \eps^2 + \eps^3 + \eps^4 + \O(\eps^5)\)\cr
&=&  \tr R \frac{1}{1-\eps}
\eea
Taking $\eps = 1$ gives infinity unless $R = 0$. The condition (\ref{oned}) is satisfied when $R$ is a $1\times 1$ matrix corresponding to $d=1$ in which case $R=0$, and yet $\tr K=0/0$ can be nonzero and finite but we can not compute it by inverting Gauss-Codazzi when $d=1$.

We can also make the ansatz more restrictive, such that we may find a closed formula. Let us make the ansatz
\ben
K_{\mu\nu} &=& R_{\mu\nu} f(R,R_{\kappa\tau})\cr
K &=& R f(R,R_{\kappa\tau})\label{ans}
\een
Inserting this ansatz into the Gauss-Codazzi equation
\bea
R &=& K^2 - K_{\mu\nu} K^{\mu\nu}
\eea
gives us
\bea
f(R,R_{\kappa\tau}) &=& \sqrt{R}{R^2-R_{\kappa\tau} R^{\kappa\tau}}
\eea
We then get the solution
\bea
K_{\mu\nu} &=& \frac{R_{\mu\nu} \sqrt{R}}{\sqrt{R^2 - R_{\kappa\tau} R^{\kappa\tau}}}\cr
K &=& \frac{R^{3/2}}{\sqrt{R^2-R_{\kappa\tau} R^{\kappa\tau}}}
\eea
This solution was found in a different way in \cite{Kraus:1999di}. But it is important to note that the ansatz (\ref{ans}) is not general, so this formula does no work for general boundaries, but it works for boundaries $\mb{R}^k \times S^{d-k}$\cite{Kraus:1999di}. We can see how this formula deviates from (\ref{EX}) by expanding it out in powers of $1/d$,
\bea
K &=& R + \frac{R_{\mu\nu}R^{\mu\nu}}{R} + \frac{\(R_{\mu\nu}R^{\mu\nu}\)^2}{R^3} + ...
\eea
We now see that only the first two terms in this $1/d$ expansion agree with (\ref{EX}).

\subsection*{Note added}
After the first version of this paper appeared on arxiv, we were informed that CFT's on $S^k \times H^{d-k}$ and their gravity duals had already been studied recently in \cite{Rodriguez-Gomez:2017kxf}. However, our results disagree with \cite{Rodriguez-Gomez:2017kxf} in many places but for the finite term when $k$ is odd (and $d$ is odd) our results agree. 

\section*{Acknowledgements}
This work was supported in part by NRF Grant 2017R1A2B4003095.

\appendix
\section{Conventions for the Riemann tensor}
We define
\bea
[D_{\mu},D_{\nu}] v_{\lambda} &=& R_{\mu\nu\lambda}{}^{\rho} v_{\rho}\cr
D_{\mu} w_{\nu\lambda} &=& \partial_{\mu} w_{\nu\lambda} - \Gamma_{\mu\nu}^{\tau} w_{\tau\lambda} - \Gamma_{\mu\lambda}^{\tau} w_{\nu\tau}
\eea
Then
\bea
R_{\mu\nu\lambda}{}^{\kappa} &=& - \partial_{\mu} \Gamma_{\nu\lambda}^{\kappa} + \Gamma_{\mu\lambda}^{\tau} \Gamma_{\nu\tau}^{\kappa} - (\mu\leftrightarrow \nu)
\eea
We then define $R_{\mu\nu} = R_{\mu\lambda\nu}{}^{\lambda}$. With these conventions, spheres have postive scalar curvature.

\section{The integral $J_{d,k}$}\label{integral}
In this appendix we compute the integral
\bea
J_{d,k} &:=& \int_0^{\rho_0} d\rho \sinh^k \rho \cosh^{d-k} \rho 
\eea
We define 
\bea
r &=& \cosh\rho
\eea
Then 
\bea
J_{d,k} &=& \int_1^{r_0} dr \(r^2-1\)^{\frac{k-1}{2}} r^{d-k}
\eea
Next, if we put 
\bea
r &=& \frac{1}{\sqrt{t}}
\eea
then 
\bea
J_{d,k} &=& \frac{1}{2} \int_{1/r_0^2}^1 dt (1-t)^{\frac{k-1}{2}} t^{- \frac{d}{2} - 1}
\eea
If we think on analytic continuation in $d$, then we may now assume that $d<1$ and define the resulting integral for positive integer $d$ by analytic continuation. We may decompose the integral as
\bea
\int_{1/r_0^2}^1 &=& - \int_0^{1/r_0} + \int_0^1
\eea
and get
\bea
J_{d,k} &=& \frac{1}{2} B\(-\frac{d}{2},\frac{k+1}{2}\) - \frac{1}{2} \int_0^{1/r_0} dt (1-t)^{\frac{k-1}{2}} t^{- \frac{d}{2} -1}
\eea
The second integral is evaluated to 
\bea
- \frac{1}{2} \int_0^{1/r_0} dt (1-t)^{\frac{k-1}{2}} t^{- \frac{d}{2} - 1} &=& \frac{r_0^{d}}{d}  {}_2 F_1\(-\frac{d}{2},\frac{1-k}{2},1-\frac{d}{2},\frac{1}{r_0^2}\)
\eea
for $d<1$. By applying analytic continuation in $d$ we may then define the integral on the left-hand side by the the expression on the right-hand side for $d\geq 1$ where the integral on the left side diverges. From these considerations it becomes clear that all divergences sit in this integral, which has no finite term, and the finite term is given by the Euler beta function. 

This computation breaks down when $d$ and $k$ are both even, in which case both the beta function and the hypergeometric function become infinite. The hypergeometric function at even $d$ has the expansion,
\bea
{}_2 F_1\(-\frac{d}{2},\frac{1-k}{2},1-\frac{d}{2},\frac{1}{r_0^2}\) &=& \sum_{n=0}^{d/2} (-1)^n \binom{d/2}{n} \frac{\(\frac{1-k}{2}\)_n}{\(1-\frac{d}{2}\)_n} \frac{1}{r_0^{2n}}
\eea
where $(x)_n$ denotes the Pochhammer symbol. But this expansion hits a potential singularity at $n = d/2$ when $d$ is even, where $\(1-\frac{d}{2}\)_n = 0$. The singularity is removable when $k$ is odd, but not when $k$ is even. We would like to suggest that we shall remove the point $n=d/2$ from the sum whenever $d$ is even, both for even and odd $k$, and define a regularized version as
\bea
{}_2 \t F_1\(-\frac{d}{2},\frac{1-k}{2},1-\frac{d}{2},\frac{1}{r_0^2}\) &:=& \sum_{n=0}^{d/2-1} (-1)^n \binom{d/2}{n} \frac{\(\frac{1-k}{2}\)_n}{\(1-\frac{d}{2}\)_n} \frac{1}{r_0^{2n}}
\eea
which is finite when $d$ and $k$ are both even. But we need to add back the point that we removed, which will give rise to the log-divergent term. We thus claim that the final result when $d$ is even is given by
\ben
J_{d,k} &=& \frac{(-1)^{d/2}}{d} \frac{\(1+(-1)^k\)}{B\(\frac{d}{2},\frac{1+k-d}{2}\)} \rho_0 + \frac{r_0^d}{d} {}_2 \t F_1\(-\frac{d}{2},\frac{1-k}{2},1-\frac{d}{2},\frac{1}{r_0^2}\) + C_{d,k}\label{Jconj}
\een
where $C_{d,k}$ is a constant. Although the formula (\ref{Jconj}) was not rigorously derived, we have confirmed its correctness by comparing it with the expanded expression for $J_{d,k}$ 
\bea
J_{d,k} &=& \frac{1}{2^d} \sum_{p=0}^{d} \sum_{m=0}^k \binom{k}{m} \binom{d-k}{p-m} (-1)^m \int_0^{\rho_0} d\rho e^{\rho(d-2p)}
\eea
up to $d=6$ for various $k$ and found agreement, where the constant term is given by
\bea
C_{d,k} &=& - \frac{1}{2^p} \sum_{{p\neq d/2},p=0}^{d} \sum_{m=0}^k \binom{k}{m} \binom{d-k}{p-m} \frac{(-1)^m}{d-2p} 
\eea
By performing some numerical computations it seems that $C_{d,k} = 0$ for all even $k$ (although we have no proof of that), but it is nonvanishing for odd $k$. For instance $C_{6,1} = 22/3$. This constant does not seem to have much physical significance. After we cancel all the powelaw divergences by a renormalization procedure, we are left with a term proportional to $\rho_0$ (which is the log-divergence if we write $\rho_0 = \ln(1/\eps)$) plus this constant. This constant can be absorbed into $\rho_0$ by making a constant shift of $\rho_0$. But if one can match two renormalization procedures on both sides of the AdS-CFT duality, then this constant should probably match on both sides too.

\section{Mass of AdS by using truncated counterterm}\label{mass}
Here we reproduce some old results that have appeared in the literature. We consider the expression for the bulk mass when we foliate AdS$_{d+1}$ with $\mb{R} \times S^{d-1}$ as in (\ref{spherefol}),
\bea
M_{bulk} &=&  - \frac{\Vol(S^{d-1})}{8\pi G} (d-1) \sinh^{d-2} \rho_0 \cosh^2 \rho_0
\eea
Let us express this in terms of $r = \sinh\rho_0$ and $N = \cosh\rho_0 = \sqrt{r^2+1}$ as
\bea
M_{bulk} &=&  - \frac{\Vol(S^{d-1})}{8\pi G} (d-1) N r^{d-2} \sqrt{r^2+1}\cr
&=& - \frac{\Vol(S^{d-1})}{8\pi G} (d-1) N r^{d-1} \sqrt{1+\frac{1}{r^2}}\cr
\eea
We expand the square root
\bea
\sqrt{1+\frac{1}{r^2}} &=& \sum_{k=0}^{\infty} \frac{c_k}{r^{2k}}
\eea
where 
\bea
c_k &=& \frac{(-1)^{k+1} \Gamma\(k-\frac{1}{2}\)}{2\sqrt{\pi}\Gamma\(k+1\)}
\eea
Then the Casimir energy is identified as the term proportional to $1/r$, which corresponds to the term $k = d/2$ in the series expansion. This is the leading term that we find if we use a truncation of the counterterm. We get
\bea
M_{bulk,k=d/2} &=&  - \frac{\l^d \Vol(S^{d-1})}{8\pi G} (d-1) \frac{N}{r} c_{d/2}
\eea
We will take $r$ large and neglect $1/r$ corrections, in which case we get
\bea
M_{bulk,k=d/2} &=& - \frac{\l^d \Vol(S^{d-1})}{8\pi G} (d-1) c_{d/2}
\eea
We note that 
\bea
(d-1) c_{d/2} &=& (-1)^{d/2+1} \frac{\[(d-1)!!\]^2}{d!}
\eea
and we arrive at 
\bea
M_{bulk,k=d/2} &=& \frac{(-1)^{d/2} \Vol(S^{d-1})}{8\pi G} \frac{\[(d-1)!!\]^2}{d!}
\eea
This is the result that was obtained in \cite{Emparan:1999pm} and reproduced in
\cite{Olea:2006vd} using the Kounterterm method. This result was however first obtained for the special case when $d=4$ in \cite{Balasubramanian:1999re} where it yields the result
\bea
M_{bulk,k=2,d=4} &=& \frac{3\pi}{32 G}
\eea
Using the AdS/CFT relation $1/G = 2N^2/\pi$ it becomes $M_{bulk,k=2,d=4} = 3N^2/16$ which matches with the Casimir energy of the dual $N=4$ SYM with gauge group $SU(N)$ living on the $\mb{R}\times S^3$ boundary. 

This kind of match does not generalize to AdS$_7$ where we find 
\bea
M_{bulk,k=3,d=6} &=& - \frac{5\pi^2}{128 G}
\eea
Using the AdS/CFT relation $\pi^4/G = 32 N^3$ we get $M_{bulk,k=3,d=6} = - 5 N^3/(4\pi^2)$. But the Casimir energy for M5 brane of type $A_{N-1}$ gauge algebra on $\mb{R} \times S^5$ is given by $E = - N^3/6$ at the supersymmetric point \cite{Bak:2019lye}. So here there is a mismatch by a factor of $4\pi^2/30 = 1.316...$.

\section{The boundary component $\eta=\eta_0$}\label{other}
We have been largely ignorant about the boundary component $\eta=\eta_0$, and we have not understood what happens there. Somehow it seems we get the right results by simply discarding counterterms from this boundary component, but we do not understand whether this is a legitimate procedure. Perhaps for later reference we list the curvature components on this boundary component here. The metric on this boundary is
\bea
ds^2 &=& d\rho^2 + \sinh^2\rho d\Omega_k^2 + \cosh^2\rho d\Omega_{d-k-1}^2
\eea
where we let $d\Omega_k^2 = G_{ij} dx^i dx^j$ and $d\Omega_{d-k-1}^2 = G_{ab} dx^a dx^b$ and $ds^2 = g_{MN} dx^M dx^N$ for $x^M = (\rho,x^i,x^a)$. Then we have 
\bea
R_{ijk}{}^l &=& - \(g_{ik} \delta_j^l - g_{jk} \delta_i^l\)\cr
R_{abc}{}^d &=& - \(g_{ac} \delta_b^d - g_{bc} \delta_a^d\) \(1-\frac{1}{\cosh^2\rho} \frac{\cosh^2\eta_0}{\sinh^2\eta_0}\)\cr
R_{aib}{}^j &=& - \delta_i^j g_{ab}\cr
R_{iaj}{}^b &=& - \delta_a^b g_{ij}\cr
R_{\rho i \rho}{}^j &=& - \delta_i^j\cr
R_{a\rho b}{}^{\rho} &=& - g_{ab}\cr
R_{i \rho j}{}^{\rho} &=& - g_{ij}
\eea
From this we can see that this boundary at $\eta=\eta_0$ is that of unit $H^d$ when $R_{abc}{}^d = 0$, which happens when the index range for $a$ is sufficiently small such that we foliate $H^{d+1}$ with either $S^{d-1}\times H^1$ or with $S^{d-2}\times H^2$, and for other foliations we have an infinitesimal deviation away from $H^d$ that is caused by the components $R_{abc}{}^d$.

\section*{Figures}\label{figures}

\begin{figure}[h!]
\begin{center}
\includegraphics[height=2.5in]{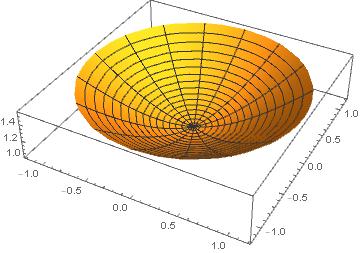}
\end{center}
\caption{Hyperbolic space with a boundary circle.}
\end{figure}

\begin{figure}[h!]
\begin{center}
\includegraphics[height=2.5in]{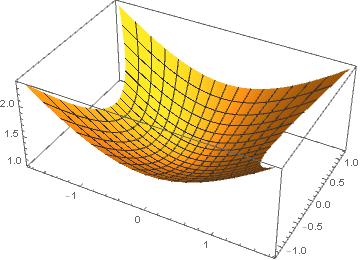}
\caption{Hyperbolic space with a boundary with four corners.}  
\end{center}
\end{figure}

\end{document}